\documentclass[twocolumn]{aastex631}

\usepackage{amsmath} 

\shorttitle{AASTeX v6.3.1 Sample article}
\shortauthors{Zeng et al.}
\graphicspath{{./}{figures/}}

\defcitealias{paperiii}{Paper III}

\begin{document}

\title{Deep Submillimeter and Radio Observations in the SSA22 Field. IV. Spectral Energy Distributions, Star Formation Histories, and the Infrared-Radio Correlation of the 850 $\micron$-selected SMGs}

\author[0009-0000-6036-7320]{Xin Zeng}
\affiliation{Purple Mountain Observatory, Chinese Academy of Sciences, Nanjing 210023, China}
\affiliation{School of Astronomy and Space Sciences, University of Science and Technology of China, Hefei 230026, China}

\author[0000-0003-3139-2724]{Yiping Ao}
\correspondingauthor{Yiping Ao} 
\email{ypao@pmo.ac.cn}
\affiliation{Purple Mountain Observatory, Chinese Academy of Sciences, Nanjing 210023, China}
\affiliation{School of Astronomy and Space Sciences, University of Science and Technology of China, Hefei 230026, China}

\author[0000-0003-4985-0201]{Ken Mawatari}
\affiliation{Waseda Research Institute for Science and Engineering, Faculty of Science and Engineering, Waseda University, 3-4-1 Okubo, Shinjuku, Tokyo 169-8555, Japan}
\affiliation{Department of Pure and Applied Physics, School of Advanced Science and Engineering, Faculty of Science and Engineering, Waseda University, 3-4-1 Okubo, Shinjuku, Tokyo 169-8555, Japan}

\author[0000-0003-1937-0573]{Hideki Umehata}
\affiliation{Department of Physics, Graduate School of Science, Nagoya University, Furocho, Chikusa, Nagoya 464-8602, Japan}
\affiliation{Institute for Advanced Research, Nagoya University, Furocho, Chikusa, Nagoya 464-8602, Japan}



\begin{abstract}
We analyze the spectral energy distributions (SEDs), star formation histories (SFHs), and infrared-radio correlation (IRRC) of 221 850 $\micron$-selected submillimeter galaxies (SMGs) in the SSA22 deep field. 
The average SEDs reveal that SMGs have cooler dust temperatures than local ULIRGs or starburst systems, and that higher-redshift SMGs tend to be more luminous and exhibit warmer dust.
The median mass-weighted age is 567 Myr. 
Most galaxies in our sample began forming $\sim$ 1.68 Gyr after the Big Bang, entered the ‘SMG phase’ after $\sim$ 1 Gyr of evolution—when they are predominantly observed—and largely transitioned out of the ‘SMG phase’ to become quiescent within an additional $\sim$ 0.2 Gyr.
A subset of massive galaxies shows rapid early assembly with high star formation efficiencies ($\sim$0.2-0.8). 
The majority of SMGs reside at the high-mass end of the star-forming main sequence, with a characteristic stellar mass of $M_{\text{star}} \sim 10^{11}$ M$_\odot$, above which galaxies are predominantly either on the main sequence or already quenched.
We observe a downsizing trend: more massive galaxies tend to “mature” earlier, completing their major episodes of star formation at higher redshifts compared to lower-mass systems.
Our sample contributes $\sim$ 21\% (28\%) to the cosmic star formation rate density (stellar mass density), including the overdensity, with its relative contribution peaking at 50-60\% in the redshift range $z=2.5-3.5$. 
%
The median infrared-radio correlation parameter $q_\text{IR}$ is 2.37, evolving as $(1+z)^{-0.11}$, likely due to AGN contributions at high redshift and intrinsic differences between low- and high-redshift populations.

\end{abstract}




\section{Introduction} 
\label{sec:intro}

Massive galaxies at high redshift serve as key probes for tracing the history of cosmic star formation. Following the epoch of cosmic reionization, starbursts embedded in dense clouds gradually became the dominant component of cosmic star formation, contributing approximately 50-80\% of the star formation rate density (SFRD) \citep{hughes1998, zavala2021}. 
We focus on a key population of such systems: SMGs—high-redshift, massive, dust-obscured star-forming galaxies. Thanks to the negative $k$-correction at (sub)millimeter wavelengths, SMGs provide a uniquely powerful window into the early universe.
They not only constitute a major component of cosmic star formation at early epochs but also offer critical insights into fundamental astrophysical questions: When and how did galaxies and stars first form? What are their evolutionary pathways? How was the universe enriched with metals over time? Moreover, SMGs serve as vital laboratories for studying galaxy assembly, dynamical evolution, and the emergence of large-scale structure.

SMGs are ultra-luminous, dust-enshrouded star-forming galaxies, widely regarded as progenitors of local ellipticals \citep{genzel2003, simpson2014, simpson2017, lim2020a, zhang2022}. 
%
Sources selected at 850 $\micron$ with flux densities $\gtrsim$ 1 mJy—having a median of $\sim$ 4 mJy—exhibit star formation rates (SFRs) ranging from tens to thousands of solar masses per year, with stellar masses typically in the range of $\sim 10^{10-12}$ M$_\odot$ \citep{swinbank2014, cunha2015, dud2020}.
They reside in massive dark matter halos of $\gtrsim 10^{13} h^{-1}$ M$_\odot$ \citep{chen2016, an2019, lim2020b} and are frequently found in association with cosmic large-scale structures \citep{dressler1980, tamura2009, battaia2018, umehata2019, zhang2024, zhang2025}.

Approximately half of the energy from cosmic star formation is channeled through the cosmic far-infrared background \citep{dole2006}, to which SMGs contribute significantly \citep{geach2017}. However, single-dish submillimeter/millimeter surveys are limited by confusion noise and sensitivity, typically resolving only 10-40\% of the cosmic infrared background \citep{wang2017, gao2024}. Next-generation facilities such as the 15-meter Xue-shan-mu-chang SubMillimeter Telescope (XSMT-15m; \citealp{xsmt2025}) are expected to substantially improve constraints on the background radiation and cosmic SFRD.

In the local universe, ultra-luminous infrared galaxies (ULIRGs) are almost exclusively merger-driven systems \citep{sanders1988, lonsdale2006}. By analogy, it was long assumed that high-redshift SMGs are similarly triggered by gas-rich major mergers \citep{tacconi2008}, consistent with their often asymmetric morphologies, clumpy structures, and kinematically disordered gas motions \citep{zadeh2012, menendez2013}. 
However, alternative models propose that SMGs can be sustained over longer timescales through steady cold gas accretion \citep{narayanan2015, mcalpine2019}. Indeed, spatially resolved spectroscopy has revealed rotationally supported disks in many SMGs \citep{genzel2011, swinbank2011, breuck2014, barro2017b, amv2024, birkin2024}, and their optical/far-infrared morphologies are often well described by exponential disk profiles \citep{targett2013, hodge2016, fujimoto2018}. 
Recent high-resolution JWST observations further suggest that many SMGs/DSFGs may be massive, isolated disk galaxies whose intense star formation is driven by gravitational instabilities in gas-rich disks \citep{gillman2023, gillman2024, lebail2024, chan2025, hodge2025, zhang2025}. Cosmological simulations also support the possibility of SMG-like systems evolving in isolation \citep{dave2010, dave2012}.

Since their discovery, SMGs have been considered likely progenitors of local elliptical galaxies due to their extreme infrared luminosities \citep{hughes1998, eales1999, lilly1999}. In the ALMA era, dynamical mass estimates, luminosity evolution, and number densities have reinforced this view, indicating that SMGs possess the necessary properties to evolve into present-day early-type galaxies \citep{swinbank2006, simpson2014}. Compact quiescent galaxies at $z \sim$ 2 share similar stellar masses, velocity dispersions, dynamical masses, sizes, and clustering properties with SMGs \citep{toft2014}, and given the short duration of the SMG phase ($\sim$ one to few 100 Myr), these quiescent systems are widely interpreted as the ``evolutionary bridges" between SMGs and local ellipticals. However, such an evolutionary path would require significant structural transformation \citep{chen2015}.
To reconcile this, \citet{barro2013} and \citet{dekel2014} proposed an evolutionary sequence wherein early star-forming galaxies undergo compaction via violent disk instabilities or major mergers, followed by quenching through gas exhaustion, stellar feedback, and AGN activity, and finally puff up via dry minor mergers. At $z \lesssim$ 2, galaxies may instead quench via morphological quenching or halo quenching, evolving directly into extended quiescent systems—though even in this scenario, without requiring a compact precursor phase. 

In the local Universe, galaxies with stellar masses $\gtrsim 10^{10.5}$ M$_\odot$ are predominantly early-type systems \citep{kelvin2014, moffett2016, driver2022}. If we assume that SMGs have gas fractions close to unity and that all gas is eventually converted into stars, most would exceed this mass threshold by $z \sim$ 0.  
But does this necessarily mean that all SMGs evolve into massive elliptical galaxies in the local Universe?
Observational and theoretical evidence suggests a more nuanced picture. From a dynamical perspective, \citet{cappellari2011} showed that about two-thirds of local early-type galaxies (Hubble types S0 and E) are still rotation-dominated, rather than pressure-supported classical ellipticals. \citet{cappellari2016} further demonstrated that massive local galaxies with log$(M_\text{star}/M_\odot) \gtrsim$ 10.5–11.3 populate distinct structural sequences but remain largely rotation-supported; only the most massive systems (log$(M_\text{star}/M_\odot) \gtrsim$ 11.3) exhibit the kinematic and morphological properties of traditional ellipticals.
\citet{gillman2023} found that, in rest-frame optical imaging, the non-merging SMGs typically exhibits disk-like morphologies (see their Figure 7).
Simulations further indicate that even after undergoing extreme major merger events, galaxies can still gradually develop a rotating disk structure provided the merger system remain gas-rich \citep{barnes2002, springel2005, robertson2006, hopkins2009}. Even in gas-poor mergers, the accretion of cold flows from the cosmic web or the cooling of hot halo gas may still facilitate the regrowth of a disk \citep{governato2009}.  
We remain open regarding whether all SMGs will eventually evolve into elliptical galaxies. A comprehensive study of massive star-forming and quenching galaxies at both high and low redshifts will be essential to further clarify this evolutionary picture.


The concept of the “main sequence” of galaxies was first introduced by \citet{noeske2007a} based on an analysis of approximately 3000 mass-selected field galaxies in the EGS deep field, revealing a robust and tight correlation between star formation rate (SFR) and stellar mass that evolves with redshift. \citet{daddi2007} similarly discovered the SFR–$M_\text{star}$ main sequence relation in 24 $\micron$-selected galaxies. 
\citet{dave2012} proposed that this main sequence arises from a self-regulating equilibrium process between galactic star formation and the surrounding gas supply, where feedback mechanisms maintain a balance between gas accretion, star formation, and outflow processes.

During the 1970s and 1980s, a remarkably tight linear correlation between infrared and radio luminosities across several orders of magnitude was discovered in galaxies \citep{helou1985, condon1992, delhaize2017}, known as the infrared–radio correlation (IRRC), commonly characterized by the parameter $q_\text{IR}$.
Although its precise physical origin remains debated, the IRRC is generally attributed to both emissions tracing star formation activity \citep{ivison2010b}. Ultraviolet radiation from young stars is reprocessed by dust into infrared emission, while relativistic electrons (Cosmic-Ray electrons) accelerated by supernova remnants of massive stars produce synchrotron radiation \citep{condon1992}.
This correlation persists across diverse environments from galaxy clusters to field galaxies, and from the local universe to high redshifts, being observed in normal star-forming galaxies, luminous starbursts \citep{helou1985}, and even AGN \citep{moric2010}.
And the IRRC does not evolve significantly with the galaxy's distance from the main sequence \citep{magnelli2015}. 
Due to its robustness, the IRRC has become a valuable tool for identifying radio-loud AGN, estimating star formation rates from radio luminosities, and determining distances to high-redshift galaxies \citep{delhaize2017}.

The persistence of the IRRC at high redshifts presents a theoretical challenge, given the rapidly increasing energy density of the cosmic microwave background (CMB) with redshift ($U_{\text{CMB}} \propto (1 + z)^4$) \citep{murphy2009b}. Under these conditions, inverse Compton (IC) scattering and ionization losses should dominate cosmic-ray (CR) electron cooling, surpassing synchrotron radiation. This could potentially breakdown the IRRC at high redshift, or equivalently, increase $q_\text{IR}$ with redshift \citep{magnelli2015}.
However, observations indicate that IC cooling has not yet overtaken synchrotron cooling at least up to $z \sim$ 2 \citep{magnelli2015}. Our sample demonstrates that the IRRC persists at least out to $z \sim$ 4, and \citet{delhaize2017} report that galaxies continue to follow the IRRC even at higher redshifts ($z \sim$ 6).
What mechanisms preserve the IRRC at high redshifts? Theoretical models suggest that enhanced galactic magnetic fields could increase the efficiency of synchrotron cooling, compensating for the increased energy losses from IC scattering and ionization due to the stronger CMB \citep{sargent2010b, yoon2024}. However, magnetic field amplification alone may be insufficient, as it would require extremely high field strengths. Moreover, under such conditions, CR electrons primarily lose energy via ionization and bremsstrahlung rather than synchrotron emission, which would further suppress non-thermal radio emission. 
Alternative explanations include the possibility that AGN activity dominates the radio emission at high redshifts, thereby maintaining the observed IRRC \citep{murphy2009a}, and/or enhanced free-free (bremsstrahlung) emission at high redshifts supplement the total radio luminosity \citep{murphy2009a, magnelli2015, delhaize2017}. A combination of these mechanisms is likely responsible for maintaining the correlation in the early universe.

This paper is organized as follows.  In Section~\ref{sec:data}, we briefly describe the submillimeter and multi-wavelength data in the SSA22 deep field.  
In Section~\ref{sec:sed}, we present the average SED of SMGs in the SSA22 field. We examine how the SEDs vary across different physical groupings of SMGs and detail the mean emission contributions from individual physical components.
%
In Section~\ref{sec:sfh}, we present a detailed analysis of SMGs, focusing on their position relative to the star-forming main sequence, their specific star formation rates (sSFR), their lifetimes, and their contributions to the cosmic star formation rate density and stellar mass density (SMD). 
%
Finally, in Section~\ref{sec:irrc}, we investigate the infrared–radio correlation in SMGs, presenting the distribution and redshift evolution of the correlation parameter $q_\text{IR}$, as well as the relationship between submillimeter and radio flux densities.  

Throughout this work, we adopt the cosmological parameters $H_0 = 70$ km s$^{-1}$ Mpc$^{-1}$, $\Omega_{\Lambda} = 0.70$, and $\Omega_m = 0.30$. A \citet{chabrier2003} initial mass function (IMF) is assumed. Where necessary, quantities from the literature based on other IMFs have been converted to the Chabrier IMF: values derived with the \citet{salpeter1955} IMF are multiplied by 0.63 for SFR and 0.61 for stellar mass, while those based on the \citet{kroupa2003} IMF are scaled by 0.94 for SFR and 0.92 for stellar mass \citep{madau2014, lim2020a}.

\section{Data}
\label{sec:data}

We present the deepest 850 $\micron$ imaging of the SSA22 deep field to date, combining data from the SCUBA-2 Cosmology Legacy Survey \citep[S2CLS;][]{geach2017} with additional observations of the central region \citep{ao2017}. The total integration time amounts to 91 hours, covering an area of 0.34 deg$^2$, with a central sensitivity reaching $\sigma_{850} \sim$ 0.79 mJy beam$^{-1}$.
The JVLA 3 GHz observations of the central area of SSA22 field were conducted in 2015 and 2016, achieving a sensitivity of 1.5 $\mu$Jy beam$^{-1}$ (before primary beam correction) and an angular resolution of 2$\farcs$3 ×2$\farcs$0.
For detailed descriptions of the SCUBA-2 850\,$\micron$ observations and data reduction, we refer readers to Paper II of this series \citep{zeng2024}; for the radio observations and data processing, see Paper I \citep{ao2017} and \citetalias{paperiii}.


Due to the inhomogeneous coverage of multiwavelength data, we perform source identification within a 0.27 deg$^2$ area of the SSA22 field.
We ultimately identify 248 SMG candidates using four complementary identification methods based on multiwavelength data: radio (JVLA) sources, 24 $\micron$ (\textit{Spitzer}/MIPS) sources, 8 $\micron$ (\textit{Spitzer}/IRAC) sources, and optical/NIR red colors.
We then estimate their photometric redshifts and derive physical properties through spectral energy distribution (SED) fitting, employing \texttt{EAZY} \citep{brammer2008} for photo-$z$ estimation and \texttt{CIGALE} \citep{boquien2019} for full SED modeling.
After applying quality cuts based on the SED fitting results, our final sample comprises 221 robust SMGs.
For details on the multiwavelength dataset, source identification, and SED fitting methodology, we refer the reader to \citetalias{paperiii}.

\section{The Average SED of SSA22 SMGs}
\label{sec:sed}

In this section, we analyze the average SED of SMGs derived from the best-fit \texttt{CIGALE} models, compare our results with literature findings (Section~\ref{sec:sed_average}), and discuss their SED variations across different redshifts and dust masses (Section~\ref{sec:sed_bin}) along with the radiation components of SMGs (Section~\ref{sec:sed_comp}).
The average SEDs of the SSA22 SMGs are available in the supplementary material.



\begin{figure*}
    \centering
    \includegraphics[width=1.\textwidth]{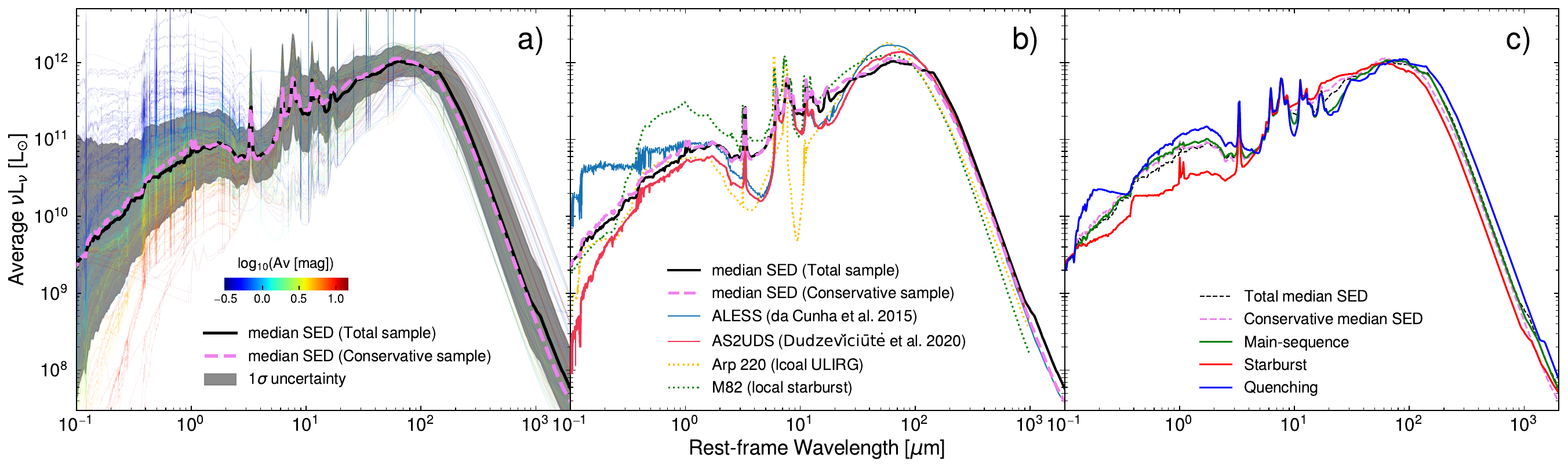}
    \caption{Panel (a): The SEDs of the 221 SMGs in the SSA22 deep field, normalized to the average infrared luminosity, and color-coded by dust attenuation $A_V$. The solid black line and gray shaded gray region represent the median SED and 1$\sigma$ dispersion, respectively.
    Panel (b): A comparison of the SSA22 SMG SEDs with those from other submillimeter surveys and local starburst galaxies and ULIRGs. Our results are consistent with the literature and exhibit colder dust temperatures than typical local ULIRGs or starbursts. However, the SED shape of SMGs more closely resembles that of local ULIRGs than that of local starbursts.
    Panel (c): SEDs of SMGs in different evolutionary stages. Quiescent SMGs display colder dust temperatures and prominent features associated with older stellar populations, while starbursting SMGs show stronger dust attenuation and warmer dust temperatures.}
    
    \label{fig:fig3-1_average_sed}
\end{figure*}


\subsection{Average SED}
\label{sec:sed_average}

Following the methodology of \citet{dud2020}, we normalized the SEDs of all 221 SMGs to a median infrared luminosity of $L_\text{IR}$ =  2.25 $\times$ 10$^{12}$ L$_{\odot}$ and computed the median flux at each wavelength to construct the average SED. 

As shown in Figure~\ref{fig:fig3-1_average_sed} image (a), the composite SED exhibits significant variations across different spectral regimes. At wavelengths $\gtrsim$ 100 $\micron$, the cold dust emission produces a broadly consistent spectral shape, though the far-infrared peak wavelength varies considerably due to differences in dust temperature, resulting in a relatively flat average spectrum with a slight peak at 60-70 $\micron$. 
Between 3-100 $\micron$, the SEDs show $\sim$ 0.5 dex scatter, primarily driven by variations in star formation-heated dust and polycyclic aromatic hydrocarbons (PAHs) emission. 
The ultraviolet-to-near-infrared (UV-to-NIR) regime displays even greater diversity (3-5 orders of magnitude), likely reflecting substantial differences in geometry, star formation activity, dust mass, and distribution—factors that complicate the optical/NIR detection of SMGs, particularly when combined with cosmological dimming effects due to $k$-correction at high redshifts\citep{dud2020}.

Color-coding the SEDs by V-band dust attenuation ($A_{V}$) reveals that SMGs with different attenuation levels can exhibit similar SED shapes. For instance, a highly obscured galaxy (due to edge-on orientation or compact dust distribution) with vigorous star formation may have an SED comparable to that of a less obscured system. This degeneracy between $A_{V}$, stellar radiation, and dust temperature underscores the diversity of SMG physical properties.

We select a conservative subsample consisting of sources with redshift $z >$ 1 and secure counterparts at VLA 3 GHz or MIPS 24 $\micron$ (see \citetalias{paperiii}). 
The average SED derived from these more reliable identifications shows a sharper far-infrared peak, and remains consistent with the average SED of the full sample. 
This demonstrates that, even if the full sample contains some level of contamination, the ensemble-averaged SED is sufficiently robust and does not alter our main conclusions.

Comparison with the ALESS \citep{cunha2015} and AS2UDS \citep{dud2020} samples (normalized to our mean infrared luminosity) shows broad agreement within 1$\sigma$ uncertainties (Figure~\ref{fig:fig3-1_average_sed} panel (b)). 
\citet{dud2020} noted that the ALESS sample lacks highly dust-obscured sources, leading to relatively moderate dust attenuation on average. This may explain why ALESS galaxies show stronger stellar emission. Similarly, we suggest that the SSA22 sample either has less dust attenuation or contains more young and old stellar populations, resulting in enhanced emission from the ultraviolet to near-infrared.
Between 3–30 $\micron$, our sources exhibit PAH emission that is 1–4 times stronger than that shown in the literature.
This may be due to a higher PAH mass fraction assumed in our models or a larger fraction of hot dust (set to $\sim$ 10\%) associated with star-forming regions. However, it is more likely attributable to insufficient mid-infrared observations leading to modeling biases.
Consequently, our average SED shows weaker cold dust emission at $\sim$ 100 $\micron$ compared to ALESS/AS2UDS. Alternatively, this difference could stem from variations in the dust models used in \texttt{CIGALE} and \texttt{MAGPHYS}, or from the greater diversity of SMGs in SSA22, which exhibit a wider range of dust temperatures. Consequently, the far-infrared peak wavelengths vary significantly, resulting in a flatter average SED in this region. At longer wavelengths, the emission from colder dust remains consistent.

Comparisons with local ULIRGs and starburst galaxies, represented by Arp220 and M82, reveal distinct features: local starbursts show stronger stellar emission ($<$ 5 $\micron$) and hotter dust continua with prominent PAH features due to the heat by nearby starburst activity (5-30 \micron) \citep{menendez2009}. ULIRGs exhibit deeper 9.7 $\micron$ silicate absorption and warmer dust temperatures than SMGs. 
Crucially, \citet{menendez2009} also pointed out that SMGs are not direct high-redshift analogs of local starbursts or ULIRGs, spatial extent of star formation region in SMGs is significantly larger than in the local starbursts or ULIRGs. However, as noted by \citet{cunha2015} and \citet{dud2020}, local ULIRGs provide better templates for high-redshift SMGs ($z \gtrsim$ 2.5) than starburst systems like M82.

Assuming a simple blackbody model, we estimate the characteristic dust temperature ($T_\text{peak}$) from the peak wavelength of the average SED using Wien's displacement law \citep{uematsu2024}. 
When using flux densities in units of [mJy] measured at discrete frequencies, the peak temperature is given by $T_{\text{peak}}$[K] = 5099/$\lambda_{\text{peak}}[\micron] $.
Arp220 and M82 exhibit emission peaks at $\sim$ 70-90 $\micron$, corresponding to $T_\text{peak}$ $\approx$ 60-70 K. In contrast, the SSA22 and AS2UDS samples exhibit cooler dust temperatures around 50 K ($\lambda_\text{peak}$ $\sim$ 100 $\micron$). The ALESS sample falls between these values, with a temperature of approximately 60 K.

This systematic trend reveals that SMGs harbor significantly cooler dust than local ULIRGs or starburst systems, aligning with \citet{symeonidis2013}'s finding of cooler dust in low-redshift (U)LIRGs compared to their local counterparts. 
\citet{simpson2017, dud2020} further demonstrate that 870 \micron-selected AS2UDS SMGs preferentially detect colder systems, with dust temperatures $\sim$ 7-8 K lower than low-redshift (U)LIRGs at comparable luminosities \citep{simpson2017}.
\citet{symeonidis2013} attribute this to increasing dust mass and/or physical extent of (U)LIRGs with redshift, a interpretation supported by \citet{menendez2009}'s finding that SMGs have more spatially extended dust distributions than local ULIRGs.

Figure~\ref{fig:fig3-1_average_sed} (c) illustrates systematic variations in the SEDs of SMGs at different evolutionary stages.
Quiescent SMGs—defined as those lying more than a factor of 3 below the star-forming main sequence (see Section~\ref{sec:sfh_ms})—exhibit cooler dust temperatures and show clear signatures of older stellar populations, including a prominent 1.6 $\micron$ stellar bump and stronger Balmer/4000\text{\AA} breaks.
Additionally, an elevation in far-ultraviolet flux is observed, likely resulting from the buildup of low-mass post-AGB stars \citep{bc03}.
In contrast, starbursting SMGs (with SFRs exceeding 3 times the main-sequence value) are characterized by stronger dust attenuation and warmer dust temperatures.


\begin{figure*}
    \centering
    \includegraphics[width=\textwidth]{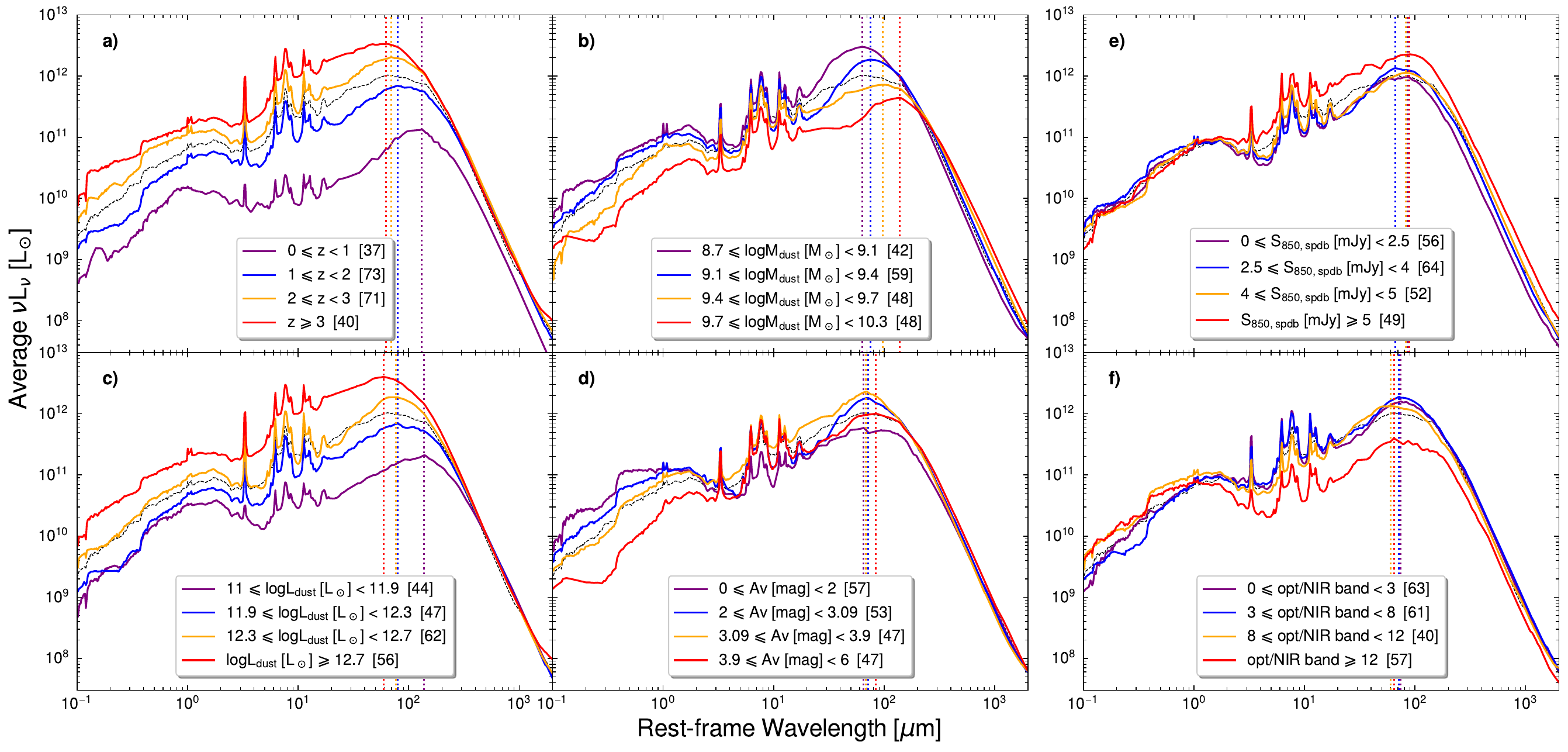}
    \caption{The median SEDs for SMGs for groups categorized by different physical or observational properties. The number of sources in each group shown within brackets in the legend. 
    The black dashed curve represents the median SED of the total sample, while vertical dotted lines indicate the peak of FIR emission for each group's SED.
    Notably, the average SEDs within each group are not normalized to their infrared luminosity. }
    
    \label{fig:fig3-2_bin_sed}
\end{figure*}


\subsection{SED Variations Across Bins}
\label{sec:sed_bin}

We grouped the SSA22 SMGs according to various physical and observational properties and computed the average SED for each group. To clearly highlight differences, the average SEDs are not normalized to infrared luminosity.

Figure~\ref{fig:fig3-2_bin_sed} panel (a) shows the average SMG SEDs in different redshift bins.
Compared to higher-redshift sources, low-redshift SMGs ($z <$ 1) exhibit younger stellar populations, with relatively bluer UV-optical emission, weaker dust attenuation and metal absorption features, and reduced thermal dust and PAH emission, resulting in fainter submillimeter fluxes and luminosities.
Figure~\ref{fig:fig3-2_bin_sed} panel (a) shows significant differences in average luminosity among SMGs at different redshifts, which may be partly due to biases introduced by the dependence of SED fitting on far-infrared (FIR) data\citep{swinbank2014, dud2020}.
As redshift increases, the total SED luminosity rises and the far-infrared peak shifts to shorter wavelengths, equally increasing characteristic dust temperature. 

For the results presented above and in the following sections, we should bear in mind several potential biases that may affect their interpretation.
First, SED fitting relies on FIR band constraints. This introduces a selection bias toward intrinsically brighter sources at high redshifts, while low-luminosity galaxies at similar redshifts may fall below the FIR detection limit. 
Consequently, some studies argue that the apparent luminosity evolution of SMGs is largely an artifact of this observational bias \citep{swinbank2014, dud2020, drew2022}.
However, other works contend that such biases alone cannot fully account for the observed trends \citep{zavala2018, lim2020a}. They suggest that infrared luminosity and dust temperature are still governed by underlying physical processes, with dust temperatures evolving with cosmic time \citep{symeonidis2013, sommovigo2022, viero2022, mitsuhashi2024}. In particular, a more fundamental correlation likely exists among infrared luminosity, dust temperature, and redshift \citep{symeonidis2013, zavala2018}.
Second, there is a significant bandpass selection effect (see also Section~\ref{sec:sfh_ms} and Section~\ref{sec:sfh_evolution}).  
The selection effect at 850 $\micron$ is fundamentally tied to dust mass. At a fixed infrared luminosity, SMGs/DSFGs with larger dust masses exhibit colder dust SEDs \citep{cunha2015}.
Consequently, an 850 $\micron$ survey with a fixed flux limit may miss sources that are warmer, more luminous in the infrared, but lower in dust mass. This implies that single-band 850 $\micron$ surveys may suffer from incompleteness.
Moreover, due to the negative $k$-correction in the submillimeter/millimeter regime, observations at longer wavelengths are more efficient at detecting high-redshift sources \citep{casey2014, zavala2014, wang2017}, resulting in samples with higher median redshifts.
Adding further complexity are uncertainties in the detection and physical property estimation of high-redshift galaxies, as well as the intrinsic diversity of SMGs themselves—in terms of dust temperature, geometry, star formation histories, and AGN contributions. These issues are beyond the scope of this study. Readers should simply be aware of these potential biases when interpreting our results.


Figure~\ref{fig:fig3-2_bin_sed} image (b) reveals that galaxies with higher dust mass have lower infrared luminosity, lower dust temperatures, and stronger submillimeter fluxes and luminosities. \citet{cunha2015} noted that at a given luminosity, SMGs with lower 870 $\micron$ fluxes tend to have higher dust temperatures, while submillimeter flux is positively correlated with dust mass \citep{dud2020}. Qualitatively, dust mass and temperature exhibit an inverse correlation, so galaxies with lower dust mass tend to have higher dust temperatures (see Section 5.4 of \citetalias{paperiii} or Figure 13 of \citealt{uematsu2024}). 
Dust distribution or optical depth (i.e., compactness) may also influence dust temperature \citep{simpson2017, dud2020}. However, since dust mass accumulation requires time, sources with lower dust mass are preferentially found at higher redshifts, contributing to their higher luminosities.
We find no clear trend in SED properties with stellar mass. SMGs with higher stellar mass tend to show stronger emission from both young and old stellar populations and redder stellar continua, but this trend is not strict. Furthermore, thermal dust, PAH, and cold dust emission do not exhibit systematic variations with stellar mass.

Figure~\ref{fig:fig3-2_bin_sed} plot (c) displays the average SEDs in different dust luminosity bins. As dust luminosity increases, the dust emission peak shifts to shorter wavelengths, thermal dust emission strengthens, and the stellar continuum becomes steeper. Due to strong $k$-corrections, submillimeter fluxes show little variation across different infrared luminosity bins.

As V-band attenuation increases (Figure~\ref{fig:fig3-2_bin_sed} (d)), the stellar continuum of SMGs becomes progressively fainter and steeper, while dust emission intensifies overall and PAH features become more prominent. However, the trends of dust temperature and submillimeter flux with $A_V$ are weak, showing only minor differences in the highest extinction bin. Unlike dust mass, $A_V$ does not strongly correlate with dust temperature, suggesting that $A_V$ is governed by multiple factors rather than a single dominant process—such as galaxy inclination, dust mass, and geometric distribution of dust. \citet{cunha2015} argued that due to dust isotropy, galaxy inclination increases extinction without significantly altering the shape of dust emission. The differences in average SEDs may arise from variations in source compactness. In galaxies with more compact dust distributions, the higher dust column density and dust located closer to the stars experience a stronger radiation field, enhancing the efficiency of converting stellar radiation into dust emission.

This paints an evolutionary picture: as intense star formation proceeds, large amounts of dust are produced and the dust column density increases, UV photons from young stars are increasingly absorbed, and dust thermal emission strengthens. Galaxies with more vigorous star formation generally exhibit stronger infrared emission.
However, at fixed infrared luminosity, an increase in dust mass leads to a larger reservoir of dust to absorb the same radiation field, resulting in a lower equilibrium dust temperature. This cooling effect, combined with higher dust mass, typically enhances emission at submillimeter wavelengths, where the flux density scales approximately as $S_{\nu} \propto M_d \cdot T_d$ in the Rayleigh–Jeans regime.


In Figure~\ref{fig:fig3-2_bin_sed} (e), we divided the SMG sample into bins according to their deblended 850 $\micron$ flux density, with approximately equal numbers of sources per bin.
The median redshifts of the groups are similar but increase slightly with flux: 1.79, 1.90, 2.03, and 2.10. Brighter SMGs tend to lie at higher redshifts (see also Section 5.1 in \citetalias{paperiii}). Aside from the brightest group, which shows significantly higher infrared luminosity, the overall SEDs across submillimeter flux bins are quite similar. \citet{cunha2015} also found similar SEDs across flux levels and suggested that redshift and negative $k$-correction may allow intrinsically similar sources to appear at different fluxes, resulting in comparable uncertainties across groups. 
Thus, the 850 $\micron$ flux alone cannot distinguish between galaxies with different physical properties within the SMG sample. 
Conversely, this implies that our average SED (Figure~\ref{fig:fig3-1_average_sed}) adequately represents the overall population of SCUBA-2 850 $\micron$–selected galaxies in the SSA22 field.
We observe that sources with fainter 850 $\micron$ flux have relatively brighter stellar luminosity and less dust attenuation (flatter stellar continuum). As noted earlier, 850 $\micron$ flux correlates positively with dust mass; hence, fainter sources have less dust and lower dust mass fractions ($M_{\text{dust}}/M_{\text{star}}$). Dust masses in the flux bins are 0.94, 1.42, 2.17, and 3.77 $\times 10^{9} M_\odot$, with corresponding dust mass fractions of $\sim$0.6\%, 0.9\%, 1.3\%, and 2.4\%. Additionally, at similar luminosities, fainter galaxies, having less dust, exhibit higher relative dust heating efficiency. As \citet{cunha2015} pointed out, there is an inverse correlation between submillimeter flux and dust temperature: lower 850 $\micron$ flux corresponds to relatively hotter dust.

In Figure~\ref{fig:fig3-2_bin_sed} (f), SMGs are grouped by the number of available optical/NIR bands. As expected, optically brighter sources (those with more available bands) show higher optical luminosity and less dust attenuation. The median $A_V$ values for sources from optically faint to bright (fewest to most bands) are 3.87, 3.35, 3.09, and 1.72 mag, with corresponding median redshifts of 2.20, 2.05, 1.90, and 1.25. 
The SEDs of the first three groups are similar, while the fourth group differs significantly. This is likely because the group with the most available bands is dominated by much lower-redshift sources, which may be primarily composed of lower-mass SMGs, resulting in notably lower infrared luminosity.
The median dust masses of the first three groups are 2.29, 2.14, and 1.52 $\times$ 10$^{9}$ M$_{\odot}$, with dust mass fractions all around $\sim$1\%. However, optically faint SMGs exhibit higher dust attenuation, suggesting more efficient mechanisms for absorbing star formation radiation—such as more compact dust distributions (leading to higher column densities and extinction) and warmer stellar radiation ﬁeld \citep{cunha2015}, more intense star formation activity \citep{magnelli2014, cowley2017}, or a combination of both.


\begin{figure*}
    \centering
    \includegraphics[width=\textwidth]{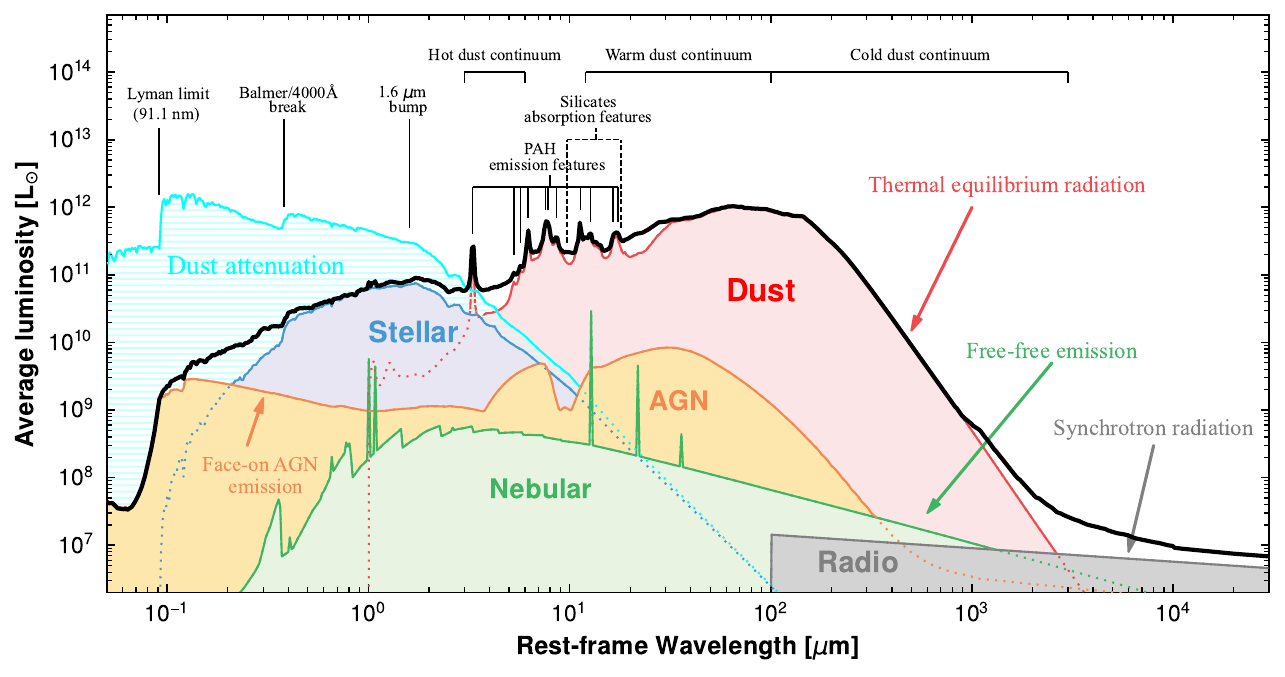}
    
    \caption{The average SED of SSA22 SMGs, along with the median contributions from various physical components. Different emission components–including stellar populations and dust features–are indicated with distinct colors and labels.}
    
    \label{fig:fig3-3_comp_sed}
\end{figure*}


\subsection{The Radiation Composition of SMGs}
\label{sec:sed_comp}

Figure~\ref{fig:fig3-3_comp_sed} exhibit the average SED of SSA22 SMGs and the average radiation composition.

In the UV-to-NIR regime, SMGs host vigorous star formation that produces abundant young stars. Their unattenuated stellar radiation includes strong UV emission, with comparable luminosity at 100 nm and 2 $\micron$. 
However, the characteristic dust-rich nature of SMGs leads to substantial attenuation, with an average V-band extinction of 3.09 magnitudes. The UV portion of the spectrum also shows significant contributions from AGN activity.

The gas-rich environment of SMGs produces prominent spectral features, including the Lyman break at 91.1 nm and Lyman series absorption shortward of 121.6 nm. 

The composite spectrum also exhibits clear Balmer/4000$\text{\AA}$ breaks - sensitive diagnostics of stellar population age, metallicity, and Lyman continuum escape fraction \citep{wilkins2023}.
The Balmer break primarily originates from A-type stars (T$_\mathrm{eff}$ $\sim$ 10000 K), where hydrogen atoms efficiently populate the second energy level, creating bound-free absorption near 3646 $\text{\AA}$ \citep{salvato2019, wilkins2023, vikaeus2024}. 
Dust attenuation enhances this feature \citep{vikaeus2024}, while its evolution tracks stellar population age: it peaks in systems dominated by 0.3-1 Gyr stars, gets filled in by strong UV continuum from younger stellar population and nebular, and smooths out in older population \citep{bc03, vikaeus2024}. 
The 4000 $\text{\AA}$ break, caused by metal absorption lines (e.g., Ca II H/K, G-band, various metal lines heavier than helium) in late-type stars \citep{bc03}, becomes more prominent in quiescent galaxies \citep{mo2010, dud2020}.

A near-universal feature in stellar spectra (except very young $\sim$ 1 Myr populations) is the 1.6 $\micron$ bump, arising from minimum H$^-$ opacity in cool stellar atmospheres. This feature proves particularly useful for photometric redshift estimation \citep{sawicki2002}.

While the origin of the 220 nm extinction bump remains debated, it has been attributed to PAH absorption \citep{fitzpatrick1999, shivaei2022}. Its absence in our SMG SEDs may result from spectral smoothing due to source diversity, or alternatively, reflect lower PAH abundances in high-redshift systems during early dust enrichment phases.

At 3 $\micron$, dust emission becomes significant and begins to dominate over the radiation from older stellar populations in SMGs. The mid-infrared emission from SMGs consists of two primary components: thermal dust continuum and PAH/molecular emission features \citep{menendez2009}.
The continuum emission at wavelengths $\gtrsim$ 12 $\micron$ originates from very small grains (VSGs; $\lesssim$ 10 nm) heated to temperatures $\lesssim$ 250 K in star-forming regions or near obscured AGN. In contrast, the $\lesssim$ 6 $\micron$ continuum arises from hotter dust ($\gtrsim$ 500 K) located in AGN or compact starburst nuclei \citep{menendez2009}. 

Between these regimes, the spectrum exhibits prominent PAH emission features excited by UV photons in star-forming regions. These nanometer-sized PAH molecules, composed of aromatic rings with attached hydrogen and trace elements \citep[e.g., Si, Mg; ][]{menendez2009}, produce characteristic vibrational bands at 3.3, 6.2, 7.7, 8.6, 11.3, 12.7, and 17 $\micron$ from C-C/C-H bond stretching and bending modes\citep{dl07}. 
The prominent absorption features at 9.7 and 18 $\micron$ arise from amorphous silicate grains\citep{dud2020}.

At longer wavelengths, $\gtrsim$ 100 $\micron$, the far-infrared emission transitions to larger, cooler dust grains. These grains reach thermal equilibrium in moderate radiation fields, emitting as modified blackbody with temperatures of 30-60 K \citep{casey2014, boquien2019}. The Rayleigh-Jeans tail at $\sim$ 100-1000 $\micron$ follows a $\nu^2$ dependence modified by the dust emissivity index (typically 1.5-1.8), resulting in an overall $\nu^{3.5-3.8}$ flux decay. The modified blackbody model has become the standard approach for characterizing far-infrared dust emission in galaxies \citep{simpson2019, dud2020, liao2024}. 
This spectral shape, combined with negative $k$-correction, produces a remarkable observational advantage in the 500-3000 $\micron$ regime: galaxies of fixed intrinsic luminosity maintain nearly constant observed flux densities across a wide redshift range (z $\sim$ 1-6) \citep{blain2002}. 
In fact, the negative $k$-correction causes galaxies observed at wavelengths longer than 1 mm to exhibit higher fluxes at higher redshifts.
These properties make (sub)millimeter observations particularly effective for detecting high-redshift dusty galaxies.

The SED turnover at $\sim$ 1-1.5 mm marks the transition where nebular emission and radio radiation begin to dominate over dust continuum. Nebular radiation primarily originates from thermal free-free emission in HII regions ionized by stellar UV photons. This bremsstrahlung process, which cools thermal electrons in unmagnetized plasmas, exhibits a characteristic flat spectrum (spectral index $\sim$ -0.1) at low frequencies ($h\nu \ll kT$), becoming steeper at higher frequencies. As noted by \citet{condon1992}, this emission becomes significant in the 30-200 GHz window where neither far-infrared nor radio continuum dominates.

The long-wavelength SED is governed by synchrotron radiation from relativistic electrons accelerated in supernova remnants of massive ($\gtrsim$ 8 M$_{\odot}$) stars \citep{condon1992}. The power-law energy distribution of these electrons produces a corresponding power-law synchrotron spectrum with typical indices of 0.3-2.0. 
Studies of star-forming galaxies reveal a median spectral index of $\alpha \sim$ 0.8 at rest-frame 1.3-10 GHz, flattening at lower frequencies due to free-free absorption and spectral aging from energy losses \citep{thomson2019, an2021, an2024}. Both the high- and low-frequency of the radio spectrum become slightly steeper with increasing stellar mass, a trend that can be explained by age-related synchrotron losses, where cosmic-ray electrons lose energy over time \citep{an2021}.
Massive stars have short lifetimes ($\lesssim$ 3 $\times$ 10$^7$ yr), and the lifetimes of relativistic electrons are also limited ($\lesssim 10^8$ yr), making radio emission a tracer of relatively recent star formation activity \citep{condon1992}. 
However, accurately observing and constraining the star formation activity in galaxies remains challenging. Although synchrotron radiation is highly directional due to the Doppler effect of relativistic electrons, the observed emission in galaxies typically originates from relatively “old” electrons ($\gtrsim 10^7$ yr), which have long since propagated far ($\gtrsim 1$ kpc) from their short-lived ($\gtrsim 10^5$ yr) supernova remnants. As a result, the original sites of electron acceleration have already disappeared, and their spatial distribution has been significantly smoothed out. Consequently, radio emission observed in galaxies appears diffuse across the entire galactic extent \citep{condon1992}.

Far-infrared–radio correlation has been observed in spiral galaxies, starburst nuclei in both galaxy cluster and blank field. This correlation is a general feature of galaxies undergoing intense star formation \citet{helou1985}. Although the exact origin of this correlation remains debated, a prevailing view holds that massive stars are the key factor \citet{condon1992}, both radio and far-infrared emissions trace star formation activity: far-infrared emission reflects dust heated by star formation, while radio emission traces synchrotron radiation from relativistic electrons in supernova remnants \citep{shim2022, zavala2018}.

AGN contribute across multiple wavebands, predominantly in the mid-infrared and radio (with additional contributions in the UV and optical for type-1 AGN).
AGN activity is frequently observed in SMGs, with reported fractions of $\sim$ 15–20\% \citep{wang2013, an2019, stach2019, uematsu2025}, brighter and more massive SMGs appear to be more likely to host AGNs \citep{wang2013, cowie2018, lim2020a}.
However, studies have shown that at least $\sim$ 85\% of the far-infrared emission in SMGs is powered by star formation activity \citep{laird2010}. In general, AGN contribute little to the total luminosity and even less to the far-infrared and millimeter emission of SMGs, typically $\lesssim$ 10\% \citep{johnson2013}. As seen in our results, these findings are well supported.

\section{The Star-Formation History}
\label{sec:sfh}

This chapter is organized according to the following logical structure:
In section~\ref{sec:sfh_ms}, we explore the relationship between SMGs and the star-forming main sequence, highlighting trends in specific star formation rate (sSFR) and the downsizing effect. 
Section~\ref{sec:sfh_assembly} reports on the stellar mass assembly histories of SMGs, including a discussion of their characteristic ages and lifetimes.   
Section~\ref{sec:sfh_contribution} quantifies the contribution of SMGs to the cosmic star formation rate density and stellar mass density (SMD).  
Finally, we presented a phenomenological picture of 850 $\micron$-selected SMG evolution based on statistical results (Section~\ref{sec:sfh_evolution}), and briefly discussed the downsizing trend as well as the reasons why submillimeter surveys typically struggle to detect SMGs at $z \lesssim$ 1.

\begin{figure*}
    \centering
    \includegraphics[width=0.9\textwidth]{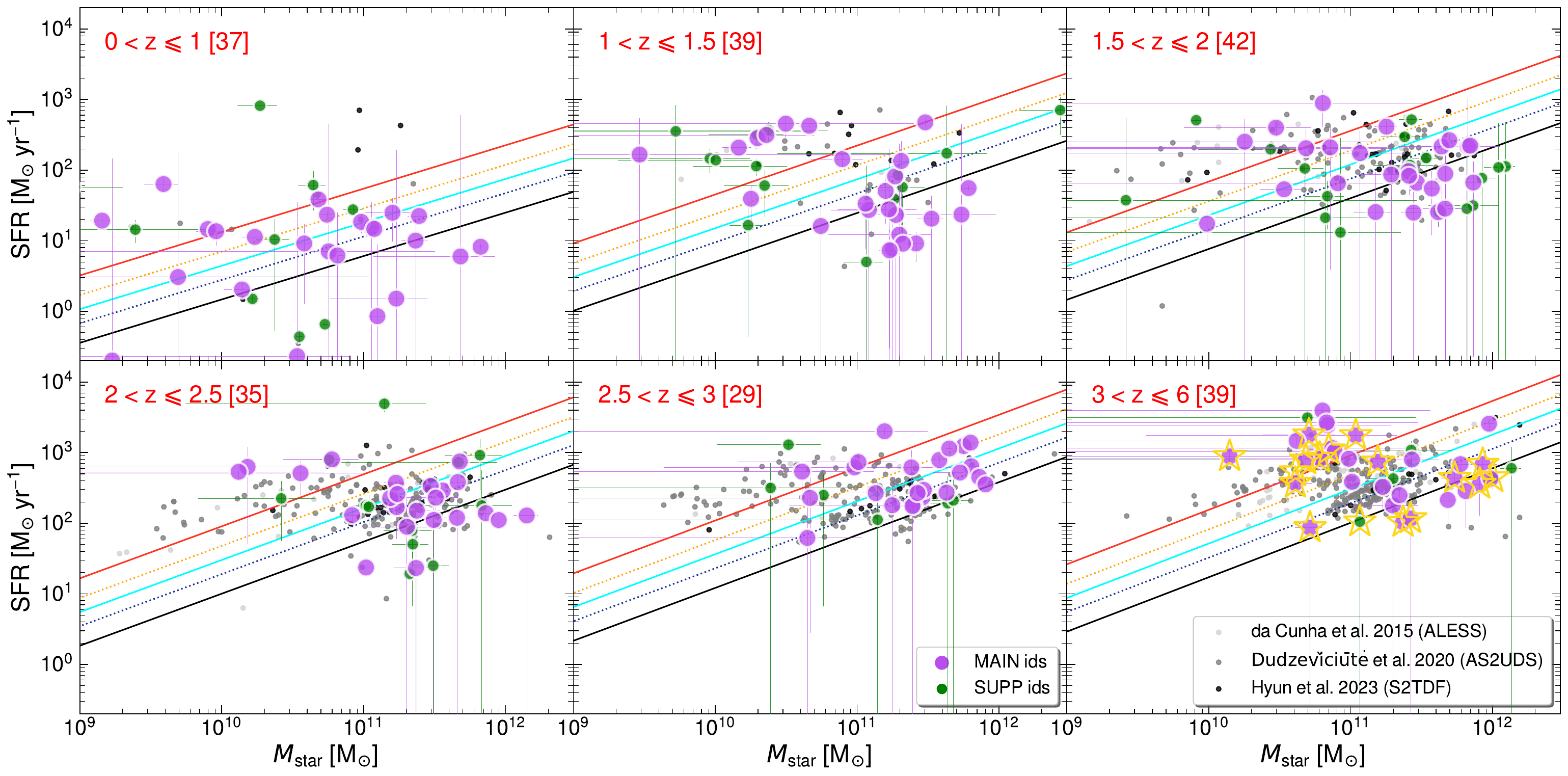}
    
    \caption{The distribution of SMGs on the SFR-$M_\text{star}$ panel. The cyan solid line denotes the main sequence relation \citep{speagle2014} at the central value of the redshift range, with yellow and blue dashed lines representing the 1$\sigma$ scatter (0.2 dex). The outermost red and black solid lines delineate the 2.5$\sigma$ dispersion; SMGs within this boundary are considered  main sequence galaxies, while those outside are classified as starburst or quiescent systems. 
    Gray dots represent SMGs from ALMA surveys \citep{cunha2015, dud2020, hyun2023}.
    It's important to note that classification actually requires calculations based on individual galaxy redshifts and masses.
    Purple and green circles represent counterparts identified through different methods (see \citetalias{paperiii} in the series), showing consistent distributions of SMGs across methods. Yellow stars indicate SMGs within the redshift range of 3.09 $\pm$ 0.1, most of which remain on the main sequence or starburst phase.}
    
    \label{fig:fig4-3_ms}
\end{figure*}

\begin{figure*}
    \centering
    \includegraphics[width=0.9\textwidth]{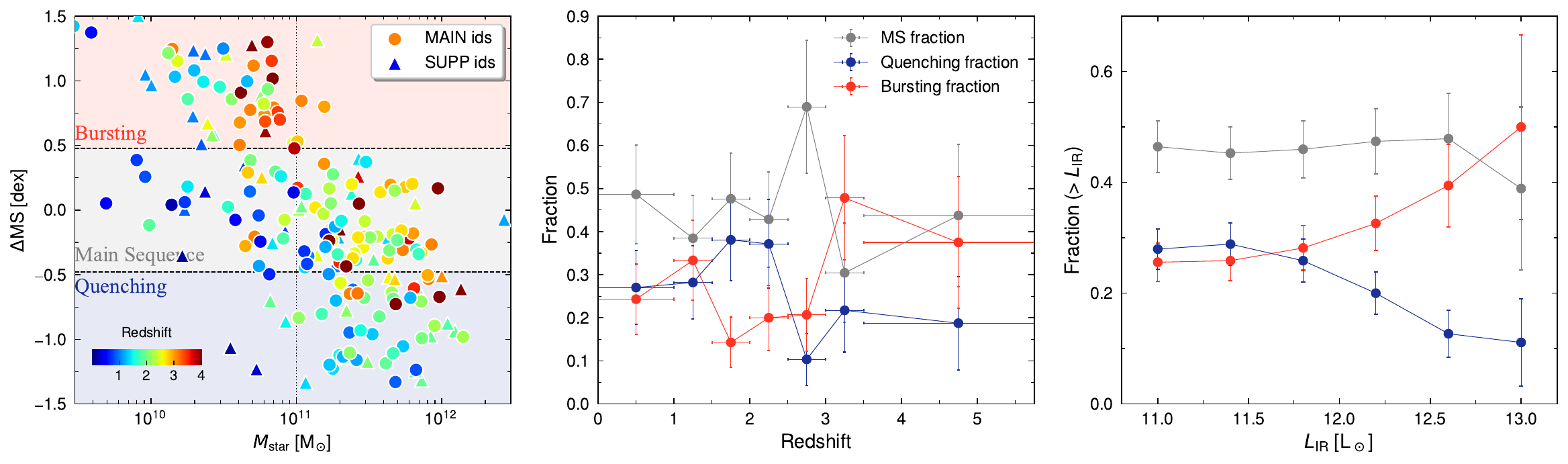}
    
    \caption{Left: The distribution of galaxies' offset ($\Delta$MS = log$_\text{10}$(SFR/SFR$_\text{MS}$) )from the main sequence against stellar mass. The horizontal dashed lines serve as a boundary between the main sequence and quiescent (or starburst) states.
    The vertical dashed line indicates a characteristic stellar mass $M_\text{star} \gtrsim 10^{11} M_\odot$; above this threshold, SMGs are predominantly quiescent or on the main sequence, while lower-mass systems tend to be starbursts or main-sequence galaxies.
    Circles and triangles denote counterparts identified through different methods (refer to \citetalias{paperiii} in the series), showing broadly consistent distributions. %
    Middle: Evolution of SMGs in different evolutionary stages with redshift. 
    Right: Fraction of SMGs in different evolutionary stages, conditioned on exceeding a given luminosity threshold.}
    
    \label{fig:fig4-4_deltams}
\end{figure*}

\subsection{Main-sequence of SMGs}
\label{sec:sfh_ms}


Figure~\ref{fig:fig4-3_ms} presents the SFR–$M_\text{star}$ distribution for SMGs across different redshift slices. 
We plot \citet{speagle2014}'s main-sequence relation as a solid cyan line in the figure:
\begin{equation}
  \log_{\rm 10}\text{SFR} = (0.84 - 0.026t) \times \log_{\rm 10}M_\text{star} - (6.51 - 0.11t),
  \label{eq:ms}
\end{equation}
where $t$ is the age of the Universe in Gyr. 
The dotted lines indicating a 0.2 dex uncertainty, which represents the intrinsic $1\sigma$ scatter of the main sequence \citep{speagle2014}.

We define the main sequence range as the region between 1/3 and 3 times the main sequence SFR (red and black solid lines in figure~\ref{fig:fig4-3_ms}, and see also left image of figure~\ref{fig:fig4-4_deltams}), encompassing approximately $2.5\sigma$ of the intrinsic scatter. 
This means that if an SMG is located at a position three times above the main sequence (taking into account the SFR, $M_\text{star}$, and redshift of each source), we define it as a starburst system \citep{cunha2015, barrufet2020, shim2022}. 
This criterion is adopted to facilitate comparisons with the literature. We note that this threshold is not strictly standardized; some studies use a factor of two \citep{elbaz2011, aravena2020}, while others adopt a factor of four \citep{dud2020, lim2020a}.
Similarly, SMGs lying more than three times below the main sequence are classified as quenched systems.

We can see that at early epochs ($z > 2.5$), SMGs are predominantly in either the starburst or main-sequence phase.
The SSA22 protocluster is located at $z = 3.09$. Accounting for photometric redshift uncertainties, we associate sources within $z = 3.09 \pm 0.1$ with this large-scale structure and mark them with yellow stars. 
Some of these may be enhanced in activity due to environmental effects, while others could be undergoing accelerated (“over”) evolution as a result of the dense environment.
At later times ($z \lesssim 2.5$), the most massive galaxies ($M_\star > 10^{11}$ M$_\odot$) begin to quench, and the star formation activity among 850 $\micron$–selected galaxies gradually becomes dominated by lower-mass systems.

Left image of figure~\ref{fig:fig4-4_deltams} shows the offset of galaxies from the main sequence as a function of stellar mass.
We identify a “characteristic mass” scale, $M_{\star} \gtrsim 10^{11}$ M$_\odot$, above which SMGs are almost exclusively either on the main sequence or already quenched. In fact, nearly all quiescent SMGs lie above this threshold.
Moreover, we find that more massive galaxies tend to “mature” earlier—consistent with the findings of \citet{firmani2010}, who reported that the mass of galaxies in transition increases with redshift. Conversely, SMGs that undergo their starburst phase at later cosmic times tend to be less massive.
This behavior aligns with the “downsizing” paradigm \citep{cowie1996}: by $z \lesssim 2.5$, the most massive 850 $\micron$–selected SMGs have already assembled the bulk of their stellar mass, and their star formation rates decline as they transition into quiescence. Consequently, the dominant contributors to dust-obscured star formation shift toward lower-mass galaxies at later epochs.

\subsubsection{Evolutionary State of SMGs}

Next, we examine in detail the fractions of 850 $\micron$–selected SMGs in different evolutionary states and how these fractions evolve with redshift.
However, we emphasize that—despite the negative $k$-correction at 850 $\micron$, which renders our luminosity limit nearly constant ($\sim 10^{12}$ L$_\odot$) across redshifts $z \sim 1–6$ for a fixed detection threshold—the 850 $\micron$ selection itself still introduces significant biases. 
For instance, 850 $\micron$ observations preferentially select galaxies with cooler dust temperatures \citep{chen2022} (see also Section~\ref{sec:sfh_evolution}). Additionally, bandpass selection effects favor the 850 $\micron$ detection of dusty galaxies at $z \sim 2–3$, while sources at both higher and lower redshifts are often missed—particularly at very high redshifts (e.g., near the end of reionization), where dusty galaxies are intrinsically rare.
A thorough resolution of these selection effects would require comprehensive multiwavelength surveys of dusty galaxies—well beyond the scope of this work.
Readers should simply be aware of these potential biases when evaluating our conclusions.

Based on previous definition, the fraction of SMGs classified as starbursts ranges from $\sim$ 20\% to 40\% across different redshifts. On average, $\sim$ 26\% of SMGs are starbursts, in excellent agreement with results for ALESS SMGs \citep[$\sim$ 27\%;][]{cunha2015}, and slightly lower than findings for SCUBA-2 sources in the NEP deep field \citep{shim2022}. 

Our sample further suggests a trend where the fraction of starburst systems increases with 850 $\micron$ flux and infrared luminosity (the right plot of Figure~\ref{fig:fig4-4_deltams}).
\citet{barrufet2020} studied a sample of high-redshift submillimeter-bright galaxies and Herschel color-selected galaxies, which are biased toward brighter submillimeter and infrared sources. Their results show a higher starburst fraction—approximately 40\% at $z =$ 2–3, nearly double our measurement—and they report that Herschel color-selected galaxies exhibit higher starburst fractions than SCUBA-2-selected galaxies.


Approximately 28\% of SMGs fall into the quiescent category, a fraction comparable to that of starbursts.
Meanwhile, about half ($\sim$ 46\%) reside within the star-forming main sequence.
The relative fractions of these populations are illustrated in the middle panel of Figure~\ref{fig:fig4-4_deltams}.
We find that at higher redshifts ($z > 3.5$), starburst systems constitute about half of the sample, dominating the early evolution of SMGs. As redshift decreases, the intense star formation in these systems subsides between $z \sim 2.5–3.5$, and most SMGs transition into the stable main sequence phase—a period coinciding with the peak of 850 $\micron$-selected SMG observations. During this phase, the starburst fraction drops to $\sim$20\%, while the main sequence fraction rapidly increases to $\sim$70\%.
Subsequently, at $z \sim 2.5$, the main sequence fraction declines to the average level, while the quenched fraction rises sharply to $\sim$35–40\%. This epoch aligns with the predicted decline in baryon conversion efficiency within massive dark matter halos (and massive galaxies), suggesting that SMG evolution follows a pathway consistent with galaxies undergoing reduced cold gas accretion \citep{behroozi2013}.
Approximately 1–2 Gyr later, at $z \sim 1.5$, there is a secondary increase in the starburst fraction by about 30\%. However, this burst of activity quickly subsides at $z < 1$, with systems transitioning back to the main sequence.

\begin{figure*}[ht!]
    \centering
    \includegraphics[width=0.95\textwidth]{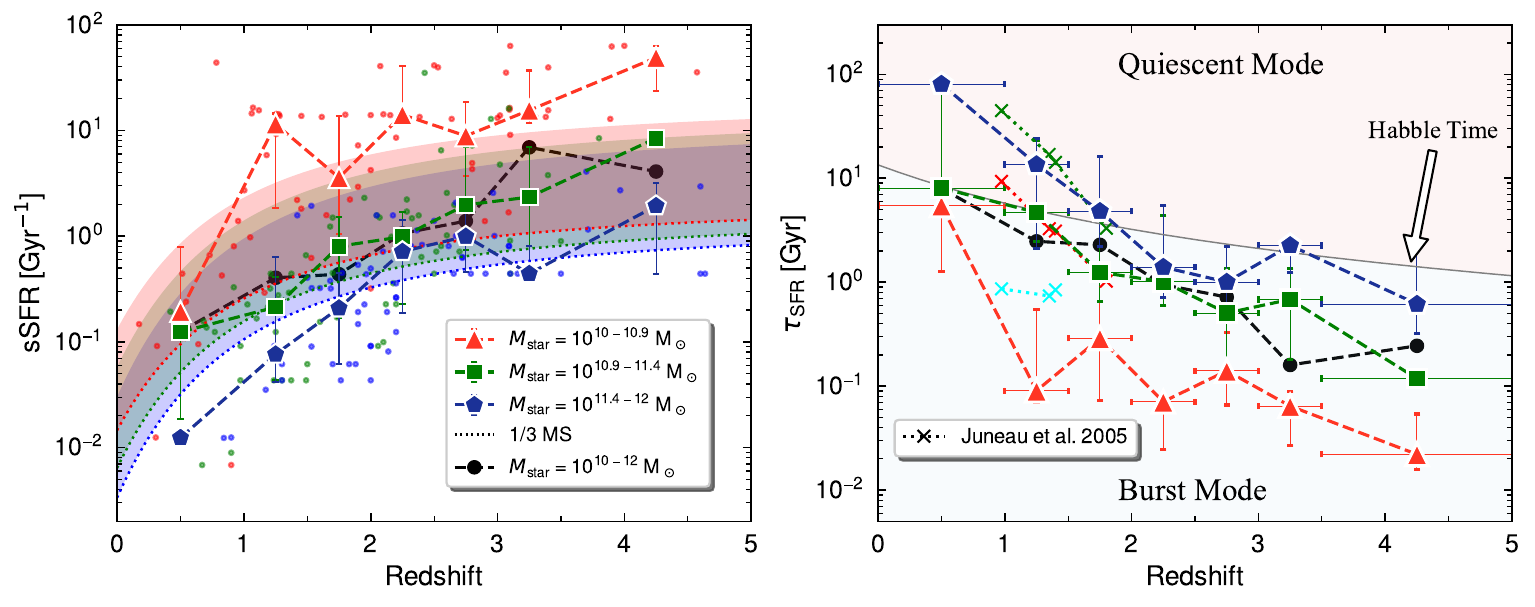}
    
    \caption{Specific star formation rate (sSFR = SFR/$M_\text{star}$) and characteristic stellar mass doubling timescale ($\tau_\mathrm{SFR} \equiv \rho_{M_\text{star}}/\rho_{SFR}$) for SMGs as a function of redshift, illustrating a clear downsizing trend: more massive galaxies complete their star formation earlier and transition to quiescence, while lower-mass systems dominate the SMG population at later times. 
    Left: sSFR versus redshift. Individual SMGs are shown as small dots, while colored symbols and corresponding shaded regions indicate the median sSFR and main-sequence scatter for different mass bins. Sources below the dotted line are evolving toward quiescence. 
    Right: Characteristic mass-doubling timescale versus redshift. The black curve represents the Hubble time at each redshift; systems above this line exhibit declining star formation activity.
    The crosses denote the results for optically and near-infrared selected galaxies from \citet{juneau2005}, with green, red, and cyan symbols representing galaxies in stellar mass bins of $10.8-11.5$, $10.2-10.8$, and $9.0-10.2$ (in $\log_{\rm 10}$ M$_\odot$), respectively.}
    
    \label{fig:fig4-5_ssfr}
\end{figure*}

\subsubsection{Specific Star Formation Rate}
\label{sec:ssfr}

The specific star formation rate (sSFR $\equiv$ SFR/$M_\text{star}$) quantifies the star formation activity per unit stellar mass in a galaxy \citep{guzman1997}, reflecting the rate at which the stellar mass fraction grows \citep{madau2014}, and is commonly used to compare the intensity of star formation across different galaxy populations.

We divide the SMG sample into three stellar mass bins: high-mass ($M_{\text{star}} = 10^{11.4-12}$ $M_\odot$), intermediate-mass ($M_{\text{star}} = 10^{10.9-11.4}$ $M_\odot$), and low-mass ($M_{\text{star}} = 10^{10-10.9}$ $M_\odot$). These bins include 90\% of the total sample, with comparable numbers of galaxies in each group. In Figure~\ref{fig:fig4-5_ssfr} left image, we plot the evolution of sSFR with redshift for the three groups, with shaded regions indicating the main sequence ranges corresponding to the median masses of each bin.

Our analysis reveals the following: (i) The sSFR of SMGs in all mass bins decreases with decreasing redshift. (ii) A clear anti-correlation exists between sSFR and stellar mass, indicating that more massive galaxies completed the bulk of their star formation earlier and now exhibit lower sSFR compared to less massive systems. This trend is consistent with findings in studies of main sequence galaxies \citep{guzman1997, brinchmann2000, juneau2005, firmani2010}. (iii) The SSA22 SMG sample clearly exhibits the ``downsizing" phenomenon: as the universe evolves to lower redshifts, star formation activity shifts from the most massive galaxies to lower-mass systems.

Taken together, we arrive at the following picture: The epoch at which baryon conversion efficiency declines in massive dark matter halos corresponds to $z \sim 2.5$ \citep{behroozi2013}. After this epoch, the most massive SMGs begin to quench, followed by intermediate-mass systems. These higher-mass groups gradually move off the main sequence after cosmic noon, reaching a passive evolutionary phase by $z \sim 1.5$.
The downsizing phenomenon leads to the quenching of the most massive galaxies \citep{miller2015}, likely due to the gradual depletion of gas reservoirs \citep{noeske2007b}. Concurrently, star formation in the universe becomes increasingly dominated by low-mass galaxies. The low-mass SMG group transitions from a starburst phase to the main sequence, reaching a stable star-forming state around $z \sim 1$.


The inverse of sSFR corresponds to the characteristic stellar mass doubling timescale \citep{madau2014}, i.e., the time required to assemble the current stellar mass at the present SFR. \citet{juneau2005} used the average timescale ($\tau_\mathrm{SFR} \equiv \rho_{M_\text{star}}/\rho_{SFR}$) to characterize the transition from burst to quiescent star formation mode across galaxy populations, defining galaxies with $\tau_\mathrm{SFR}$ shorter than the Hubble time as actively star-forming, and those with longer timescales as transitioning into quiescence. As shown in right panel of Figure~\ref{fig:fig4-5_ssfr}, applying this criterion yields conclusions consistent with our earlier classification, despite the different threshold.
Comparing with the results of \citet{juneau2005}, at similar masses, even the reddest galaxies selected in the optical and near-infrared bands quench more rapidly than SMGs. This highlights that at intermediate to low redshifts ($z < 4$), the majority of cosmic star formation occurs obscured by dust \citep{bourne2017, dunlop2017, bouwens2020, bouwens2022}, and particularly at $z \lesssim 2$, dusty star-forming galaxies contribute approximately 80\% of the total cosmic star formation activity \citep{zavala2021}.

\begin{figure*}
    \centering
    \includegraphics[width=\textwidth]{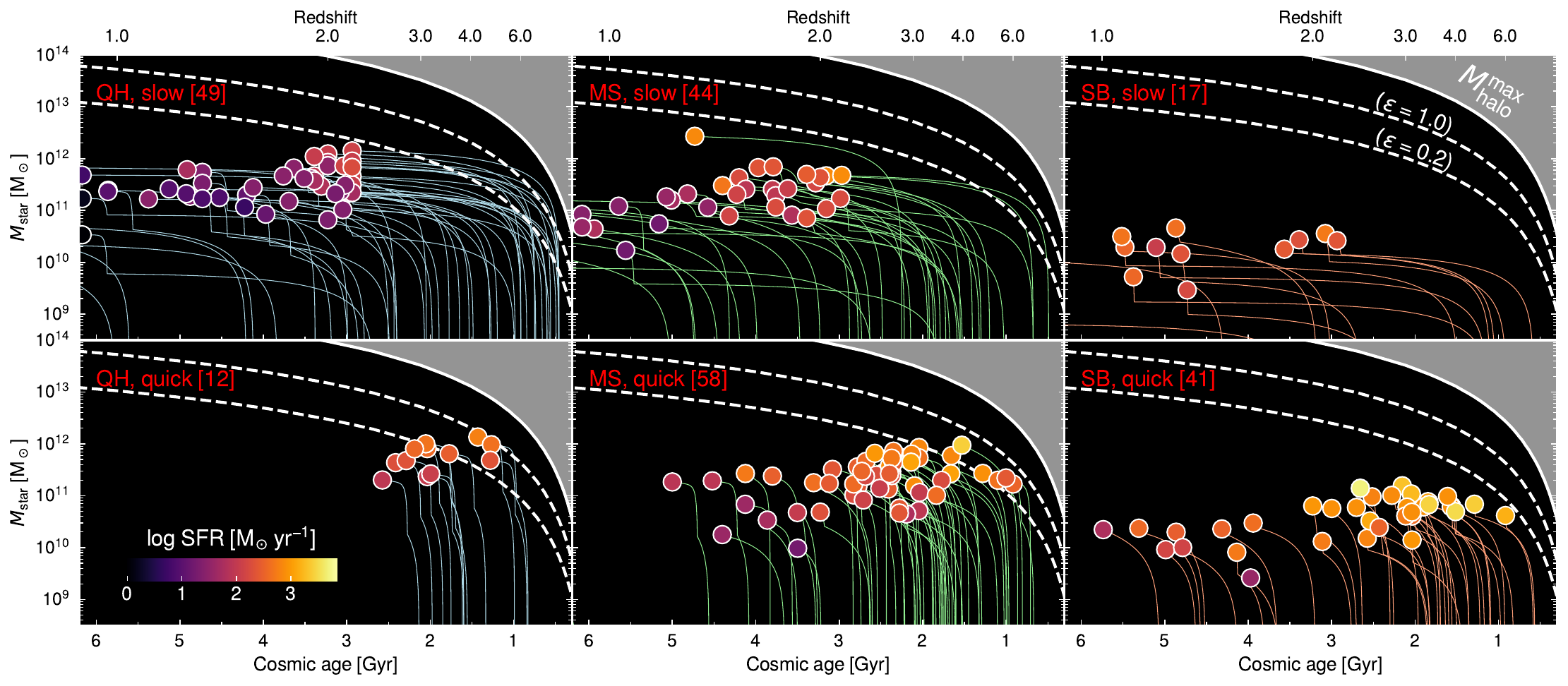}
    \caption{A visualisation of the stellar mass assembly history for each SMG, based on the best-fit star formation history from $\texttt{CIGALE}$. The red text in the upper-left corner of each subplot indicates whether the galaxy is in a quiescent phase (QH), main sequence (MS), or starburst phase (SB), as well as whether its evolutionary history is slower or faster than the median mass-weighted age $\sim$ 570 Myr. The number of SMGs in each group is given in brackets. 
    The gray shaded area represents the maximum dark matter halo mass allowed within the current survey volume.
    The dashed lines indicate the efficiency of converting baryonic matter into stars at 20\% and 100\%, assuming a baryonic fraction of 0.16 in halos.}
    
    \label{fig:fig4-1_spore}
\end{figure*}

\subsection{Stellar Mass Assembly History}
\label{sec:sfh_assembly}

Using the best-fit SEDs, we reconstruct the stellar formation histories and mass assembly trends over cosmic time for individual galaxies. 

Shown in Figure~\ref{fig:fig4-1_spore}, each ``spore" represents the assembly history of a single galaxy.
We adopt a double-exponential SFH model, which includes a dominant, extended star formation episode and a more recent burst of star formation \citep{boquien2019}.
Although the adopted SFH templates are highly simplified and likely differ from true galactic evolution, current observational techniques cannot recover exact star formation histories. Thus, these abstract models inferred from SED fitting remain valuable for studying galaxy evolution.

The median mass-weighted age $Age_\text{m}$ of the sample is 567 Myr (Table~\ref{tab:tab1_average_properties}). When SMGs are categorized by evolutionary stage into main-sequence, starburst, and quiescent subsamples (Section~\ref{sec:sfh_ms}), the median ages differ significantly, measuring 535 Myr, 374 Myr, and 1900 Myr, respectively. We find that more massive galaxies ($M_{\text{star}}$ [M$_\odot$]) $>$ 10.9) are predominantly located on the main sequence or are quiescent systems.

Furthermore, following the approach of \citet{merlin2025}, we classify the galaxies into ``fast-evolving" and ``slow-evolving" types based on whether their $Age_\text{m}$ is greater than median mass-weighted age $\sim$ 570 Myr.
The majority of quiescent SMGs ($\sim$ 80\%) have experienced prolonged evolutionary histories, averaging $\sim$ 2 Gyr. These systems typically formed at $z \gtrsim$ 3–6 and are observed around $z \sim$ 2. Most do not undergo significant secondary star formation episodes, maintaining low SFRs over extended periods. Galaxies that quench at lower redshifts show progressively less active star formation.
In contrast, ``fast-evolving" quiescent SMGs have generally undergone two intense, closely spaced starbursts, typically forming at $z \sim$ 3–4 with a median age of only 0.4 Gyr. Notably, the second burst in these systems contributes at least an order of magnitude increase in stellar mass.

Main-sequence SMGs are nearly evenly distributed between ``fast- and slow-evolving" subpopulations. ``Fast-evolving" main-sequence galaxies have ages comparable to quiescent systems (around 0.4 Gyr), but their second star formation episode is less extreme in terms of mass contribution.

Starburst SMGs are generally lower in mass. Their evolutionary tracks show rapid “upward growth”, contrasting with the more massive quiescent galaxies. The median age of starburst SMGs is $\sim$ 260 Myr.
Regardless of overall evolutionary speed, starburst galaxies generally undergo or are currently experiencing a second star formation episode, which contributes significantly to their mass assembly—particularly in systems with longer SFHs. 
This observation is consistent with the view proposed by \citet{merlin2025} that low-mass galaxies do not remain passively evolving after quenching but may reignite star formation.

Following the method of \citet{xiao2024}, we place simple constraints on the maximum galaxy mass $M_\text{gal}^\text{max} = \epsilon f_b M_\text{halo}^\text{max}$ within the survey volume in a $\Lambda$CDM framework. 
Here, $\epsilon$ is the baryon-to-star conversion efficiency, $f_b$ is the cosmic baryon fraction, and $M_\text{halo}^\text{max}$ is the maximum dark matter halo mass.
The survey coverage  ($\sim 0.27 \text{deg}^2$; Section~\ref{sec:data}) corresponds to a comoving volume of $V \sim 1.5 \times 10^7 \, \text{cMpc}^3$, assuming a conservative redshift range of 0-6. Using the Python package \texttt{hmf} \citep{murray2013}, we compute the cumulative halo mass function $n(M_\text{halo}^\text{max}, z)$. The maximum detectable halo mass at a given time satisfies $n(M_\text{halo}^\text{max}, z) \times V = 1$.
Using the cosmic baryon density $\Omega_b$ and matter density $\Omega_m$, we adopt a baryon fraction $f_b = \Omega_b / \Omega_m \simeq 0.16$ \citep{planck2018}, assumed to be constant.
We adopt $\epsilon = 0.2$ as the upper limit following common practice \citep{wechsler2018}, even though that studies suggest the conversion efficiencies range of 20-40\% \citep{behroozi2019, girelli2020, shuntov2022}.

We find that some massive galaxies underwent extremely rapid mass assembly accompanied by high star formation efficiencies (at least $\sim$ 0.2–0.8) at high redshift. These systems typically have $M_{\text{star}}$ $\gtrsim$ 10$^{\rm 11-12}$ M$_\odot$ and SFR $\gtrsim$ 200–600 M$_\odot$ yr$^{-1}$, even including those classified as quiescent. They are predominantly lie at high redshifts ($z \gtrsim 4$, i.e., within the first $\sim$ 1.5 Gyr of cosmic time). 
As noted by \citet{merlin2025}, many galaxies experienced intense bursts of star formation and rapid quenching within the first billion years, with such rare objects contributing significantly to the cosmic SFR density \citep{xiao2024}.

While the inferred efficiencies do not exceed 100\%, such high values remain puzzling. They may stem from overestimated masses or redshifts in SED fitting, or from oversimplified SFH parameterizations, which may vary substantially with environment and mass.
Alternatively, it may reflect simplistic assumptions about universal invariants—such as applying a local IMF to high-redshift galaxies, or assuming a constant baryon fraction across halo masses and redshifts \citep[even though $f_b$ depend on both;][]{girelli2020}—among other factors.

However, high-precision spectroscopic redshifts and hydrodynamical simulations confirm the existence of such systems and rule out cosmic variance \citep{xiao2024, merlin2025}. These studies also indicate that star formation in the early universe was significantly more efficient—by factors of 2–3—than at later times \citep{xiao2024, wang2025}.
\citet{dekel2023} proposed a feedback-free starbursts (FBBs) model, suggesting that high gas densities and low metallicities in massive halos at $z \sim 10$ enable highly efficient star formation. 
Key factors include short free-fall timescales ($<$ 1 Myr) in molecular clouds  and efficient gas cooling and accretion via cold streams in a halo, which avoid feedback from massive stars (e.g., winds or supernovae), while dense gas shields subsequent star formation from early generations. 
Theoretically, such galaxies are predicted to be compact with episodic SFHs \citep{dekel2023, li2024}.
As the universe evolves, increasingly extreme halo conditions are required for such efficient bursts, naturally explaining the observed decline in star formation efficiency at lower redshifts.

\begin{deluxetable*}{lccccccc}
\tablenum{1}
\tablecaption{Average properties of the SSA22 SMGs}
\label{tab:tab1_average_properties}
\tablewidth{0pt}

\tablehead{
    \colhead{Properties} &  \colhead{Total sample} & 
    \colhead{QH sample} & \colhead{MS sample} & \colhead{SB sample} & 
    \colhead{HM sample} & \colhead{IM sample} & \colhead{LM sample} \\
    \colhead{} &  \colhead{} & 
    \colhead{} & \colhead{} & \colhead{} & 
    \colhead{(10$^{11.4-12}$ M$_{\odot}$)} & \colhead{(10$^{10.9-11.4}$ M$_{\odot}$)} & \colhead{(10$^{10-10.9}$ M$_{\odot}$)} \\
    \colhead{} & \colhead{221 SMGs} &
    \colhead{61 SMGs} & \colhead{102 SMGs} & \colhead{58 SMGs} & 
    \colhead{66 SMGs} & \colhead{65 SMGs}  & \colhead{66 SMGs} }

\startdata
Age$_{\rm m}$ [Gyr] \tablenotemark{a} & 0.57$^{+1.53}_{-0.30}$  & 1.90$^{+0.23}_{-1.49}$  & 0.54$^{+1.33}_{-0.17}$  & 0.37$^{+1.08}_{-0.20}$  & 1.11$^{+0.99}_{-0.69}$  & 0.89$^{+1.21}_{-0.60}$  & 0.49$^{+1.28}_{-0.26}$  \\           
$M_{\rm star}$/SFR [Gyr] \tablenotemark{b} & 1.11$^{+6.95}_{-1.04}$ & 10.52$^{+12.59}_{-8.26}$  & 1.06$^{+2.06}_{-0.61}$    & 0.07$^{+0.04}_{-0.05}$    & 2.26$^{+9.57}_{-1.60}$    & 2.23$^{+7.35}_{-1.77}$    & 0.09$^{+2.23}_{-0.05}$    \\
100 $\times M_{\rm dust}$/SFR [Gyr] \tablenotemark{c}  & 1.31$^{+16.72}_{-1.18}$ & 6.68$^{+27.35}_{-6.08}$  & 0.97$^{+12.00}_{-0.77}$  & 0.23$^{+10.40}_{-0.17}$  & 0.98$^{+5.96}_{-0.84}$   & 1.31$^{+18.23}_{-1.13}$  & 0.52$^{+15.91}_{-0.45}$  \\ 
\hline
sSFR [Gyr$^{\rm -1}$]  & 0.90$^{+12.94}_{-0.77}$   & 0.10$^{+0.35}_{-0.05}$    & 0.94$^{+1.25}_{-0.62}$    & 14.34$^{+29.24}_{-5.23}$  & 0.44$^{+1.06}_{-0.36}$    & 0.45$^{+1.76}_{-0.34}$    & 11.05$^{+16.70}_{-10.62}$ \\
$\Delta$MS \tablenotemark{d}  & -0.16$^{+0.98}_{-0.61}$  & -0.88$^{+0.29}_{-0.33}$  & -0.15$^{+0.33}_{-0.20}$  & 0.94$^{+0.33}_{-0.27}$   & -0.38$^{+0.43}_{-0.60}$  & -0.34$^{+0.50}_{-0.52}$  & 0.67$^{+0.46}_{-0.83}$   \\
\enddata

\tablecomments{These denote, respectively, the full SMG sample; subsamples classified as quenched, main-sequence, and starburst galaxies; and SMGs divided into higher-, intermediate-, and lower-mass groups.}

\tablenotetext{a}{The mass-weighted ages of the galaxies are derived from \texttt{CIGALE}.}

\tablenotetext{b}{Characteristic star formation
timescale ($\tau_\text{sf}$) as the ratio of stellar mass to current star formation rate, $M_\text{star}$/SFR. See Section~\ref{sec:lifetime}}

\tablenotetext{c}{Characteristic gas depletion timescale, $\tau_\text{dep} = M_\text{gas}/\text{SFR} = \delta_{\text{gdr}} \times M_\text{dust}$, we adopt the commonly used gas-to-dust ratio $\delta_{\text{gdr}} \sim 100$ (Section~\ref{sec:lifetime}).}

\tablenotetext{d}{The offset of SMGs from the main sequence, defined as $\Delta$MS = $\log_{10}$(SFR/SFR$_\text{MS}$), where the main sequence relation is adopted from \citet{speagle2014}.}

\end{deluxetable*}


\begin{figure}
    \centering
    \begin{minipage}{\columnwidth}
    \includegraphics[width=\textwidth]{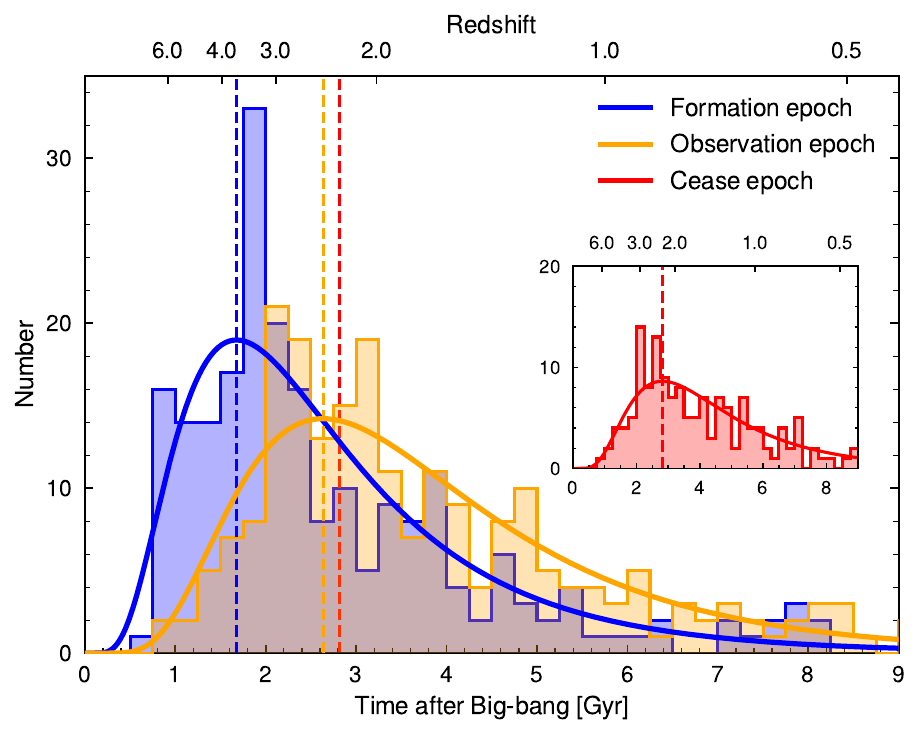}
    
    \caption{Distribution of SMGs as a function of cosmic age. Using mass-weighted ages and gas depletion timescales, we estimate the formation epoch and the time at which galaxies exit the SMG phase. 
    The solid line shows a logarithmic fit to the distribution, and the dashed line indicates the peak (mode) of the distribution, suggesting that the typical SMG in our sample exits the phase approximately 200 Myr after observation.}
    \label{fig:fig4-2_lifetime}
    
    \end{minipage}  
    
\end{figure}

\subsubsection{Age and Lifetime of SMGs}
\label{sec:lifetime}

In this section, we investigate the formation and evolutionary timescales of SMGs based on physical properties derived from SED modeling.

We estimates the ``mass-weighted age" using the best-fit SFH. Our sample shows a median mass-weighted age of 0.57 $\pm$ 0.22 Gyr, with 68\% of galaxies lying in the range 0.27–2.10 Gyr,  consistent with the average age (0.46 Gyr) of 707 ALMA SMGs in the AS2UDS survey \citep{dud2020}. 

The mass-weighted ages correlate strongly with star-forming states (Section~\ref{sec:sfh_ms}): the 102 galaxies on the main sequence have a median age of 0.54 Gyr, while the ``burst" (N=58) and ``quenching" phases (N=61) systems show 0.37 Gyr and 1.90 Gyr, respectively (Table~\ref{tab:tab1_average_properties}).
The average mass-weighted ages of SMGs in the high-mass (66 sources, 10$^{\rm 11.4\text{--}12} M_{\odot}$), intermediate-mass (65 sources, 10$^{\rm 10.9\text{--}11.4} M_{\odot}$), and low-mass (66 sources, 10$^{\rm 10\text{--}10.9} M_{\odot}$) bins are 1.11 Gyr, 0.89 Gyr and 0.49 Gyr, respectively (Table~\ref{tab:tab1_average_properties}).

We define the ``characteristic star formation timescale" $\tau_\text{sf}$ as the ratio of stellar mass to current star formation rate, $M_\text{star}$/SFR, with a median value of 1.11 $\pm$ 0.2 Gyr, matching AS2UDS results. This timescale varies by over an order of magnitude across different states: 1.06 Gyr for main-sequence, 0.07 Gyr for ``burst", and 10.52 Gyr for ``quenching" galaxies.

The gas depletion timescale, $\tau_\text{dep} = M_\text{gas}/\text{SFR}$, represents the time required to convert the gas reservoir into stars at the present SFR, assuming no replenishment of gas. 
For simplicity, we adopt the commonly used gas-to-dust ratio $\delta_{\text{gdr}} \sim 100$ \citep{swinbank2014, dud2020} to estimate the gas mass as $M_\text{gas} = \delta_{\text{gdr}} \times M_\text{dust}$. Although this approach is idealized, the gas depletion timescale remains a useful reference for estimating the duration of the SMG phase.
The gas depletion timescales of SMGs have median value of $\sim$ 1.31 Gyr in our sample, which is comparable to the $\tau_\text{sf}$ supporting the assumption that, from this perspective, observed SMGs are typically midway through their dusty phase \citep{dud2020}. The median depletion timescales for galaxies on the main sequence and in the ``burst" and ``quench" phases are 0.97 Gyr, 0.23 Gyr, and 6.68 Gyr, respectively.
It should be noted that these timescales assume idealized conditions. In reality, galaxies may accrete additional gas, and existing gas may not be fully consumed due to feedback-driven outflows or sudden quenching of star formation. Furthermore, the SFR can vary significantly over time, including episodes of starbursts or gradual decline.

Using mass-weighted ages as SMG ages and $\tau_\text{dep}$ for remaining duration, we reconstruct evolutionary timelines. Figure~\ref{fig:fig4-2_lifetime} shows the distributions of observation times, inferred onset times, and predicted end times of the SMG phase across cosmic history.
A simple median analysis suggests that the ``formation epoch" of the sample galaxies, the period when they were observed in the SMG phase, and the end of their gas depletion phase are approximately 2.21 Gyr, 3.23 Gyr, and 4.65 Gyr after the Big Bang ($z \sim$ 2.88, 2.00, 1.33), respectively.
However, the distribution exhibits a long tail and significant scatter in age space, leading to a notable difference between the median and the mode of the distribution. The peak of the observed epoch distribution occurs at an earlier time.

To better characterize the typical evolutionary path, we fit the distributions with log-normal functions and analyze the mode, $\exp(\mu - \sigma^2)$, which represents the most probable value. 
Here, $\mu$ and $\sigma$ denote the mean and standard deviation of the lognormal distribution, with the median of the lognormal distribution given by $\exp(\mu)$.
The peak of the observed epoch distribution is at 2.63 Gyr (corresponding to $z \sim 2.44$), with fitted parameters $\mu = 1.23 \pm 0.03$ and $\sigma = 0.52 \pm 0.02$. 
These massive galaxies began to form extensively around 1.68 Gyr ($z \sim 3.67$), with $\mu = 0.86 \pm 0.04$ and $\sigma = 0.58 \pm 0.03$. The end of the phase peaks at 2.82 Gyr ($z \sim 2.29$), with $\mu = 1.97 \pm 0.08$ and $\sigma = 1.23 \pm 0.06$.

From this perspective, the majority of galaxies in our sample started forming around 1.68 Gyr after the Big Bang. Approximately 1 Gyr later, they entered the ‘SMG phase’ and were observed. After an additional $\sim$ 0.2 Gyr, these galaxies largely transitioned out of the ‘SMG phase’, becoming quiescent and redder.


\subsection{The Contribution of SMGs}
\label{sec:sfh_contribution}

\begin{figure*}
    \centering
    \includegraphics[width=0.95\textwidth]{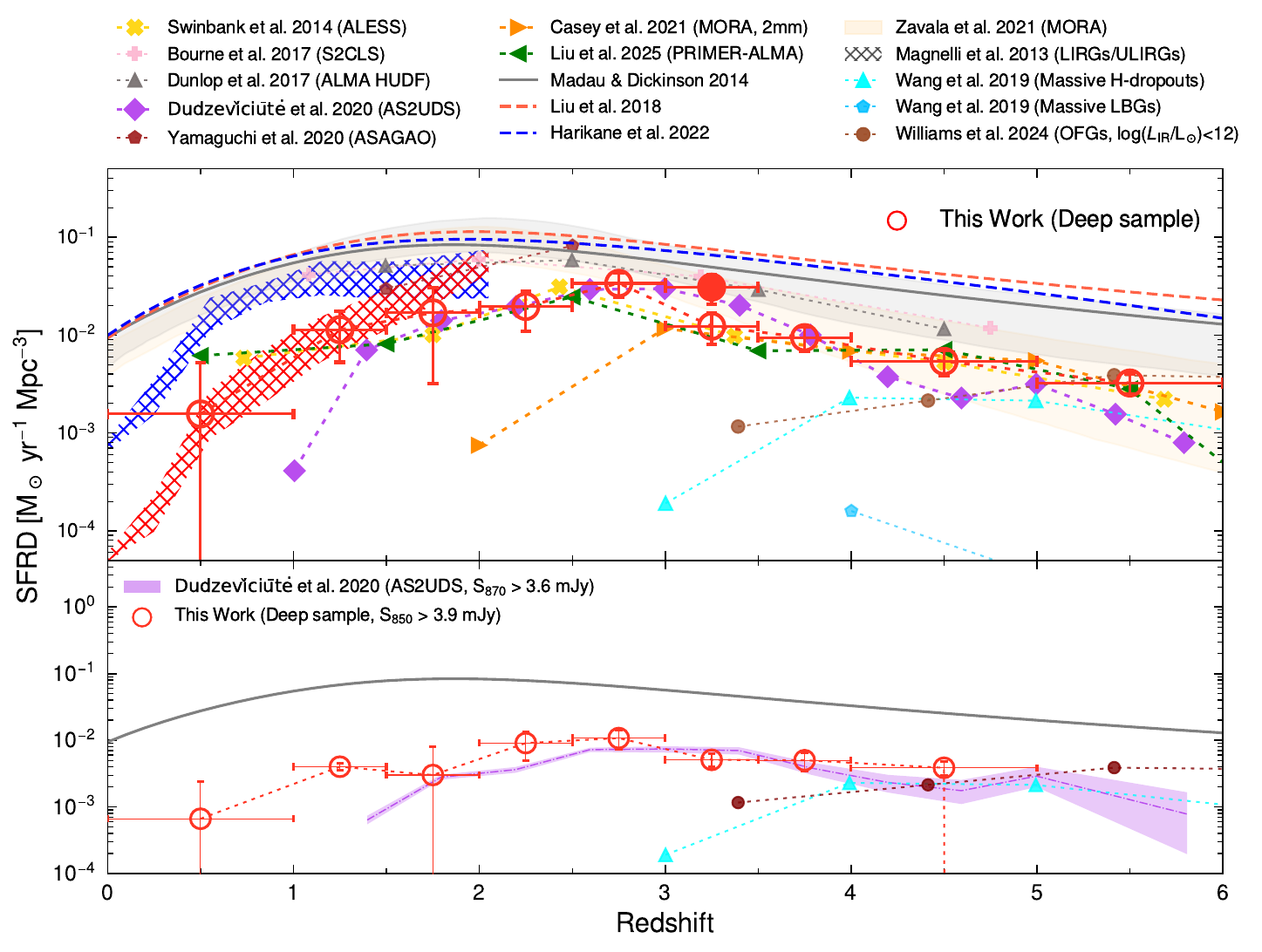}
    
    \caption{Top: Contribution of SSA22 SMGs to the cosmic star formation rate density, with a sample-averaged fraction of 23\%. Our results align with those from other large surveys, peaking within the redshift range of 2.5-3. Red filled circles include SMGs in the $z$ = 3.09 ($\pm$0.05) overdense region, highlighting the significant contribution from dense environments. 
    The red and blue grids represent the contributions from ULIRGs and LIRGs\citep{magnelli2013}, respectively, while the yellow and gray shaded regions indicate the dust-obscured star formation activity and the total SFRD inferred by \citet{zavala2021}.
    Bottom: Contribution of brighter SMGs, consistent with results from AS2UDS bright SMGs and H-dropouts. }
    
    \label{fig:fig4-6_sfrd}
\end{figure*}


\subsubsection{Contribution to the SFRD}
\label{sec:sfrd}

To quantify the contribution of SSA22 deep field SMGs to the cosmic SFRD, we first sum the SFRs of all galaxies within each redshift bin. These values are then corrected for completeness and false detection rates, and divided by the comoving volume corresponding to the survey area and redshift range (i.e., the difference in comoving volumes between the upper and lower redshift boundaries). Given that the deepest 850 $\micron$ observations ($\sigma_\text{850}$ $<$ 1 mJy beam$^{-1}$) in the central region provide the most complete multi-wavelength coverage, all statistical results discussed in this section are derived from this area. The uncertainties in SFRD estimates are based on Poisson statistics.
For consistent comparison with literature values, we apply the following conversions to SFR and stellar mass measurements originally based on different IMFs. Specifically, SFRs and stellar masses derived using a \citet{salpeter1955} IMF are multiplied by 0.63 and 0.61, respectively, while those based on a \citet{kroupa2003} IMF are scaled by 0.94 and 0.92, respectively \citep{madau2014, lim2020a}.

In addition to the results from \citet{madau2014}, we have included SFRD estimates from \citet{liu2018} and \citet{harikane2022a} in our figures. 
It is important to note that at high redshifts ($z > 4$), the SFRD curve from \citet{madau2014} is constrained by only two UV surveys. Recent dropout surveys and higher-redshift JWST observations support slightly elevated SFRD values \citep{harikane2022a, harikane2025}.
\citet{liu2018} combined direct FIR+mm observations with those from \citet{madau2014}, resulting in a total SFRD estimate that is approximately 1.5 to 1.7 times higher than that of \citet{madau2014}.
For consistency in previous literature and subsequent discussions, we reference the SFRD values from \citet{madau2014}; however, it should be noted that these values represent upper limits when compared to more recent findings.

On average, the SMG contribution in the SSA22 deep field is approximately 21\%. 
As shown in Figure~\ref{fig:fig4-6_sfrd}, our results closely match those of \citet{swinbank2014, dud2020, casey2021, liu2025}, demonstrating that counterpart identification and advanced deblending techniques successfully resolve highly confused SCUBA-2 sources into individual SMGs, pushing the 850 $\micron$ detection limit to $\sim$ 1 mJy.

By cosmic noon ($z \sim$ 2), SMGs contribute only $\sim$ 20\% of the cosmic star formation, decreasing to $<$ 5\% at $z <$ 1.
\citet{magnelli2013} also found that $z \lesssim$ 1.5, the cosmic SFRD becomes dominated by luminous infrared galaxies (LIRGs), while at later epochs it is increasingly driven by fainter sources (i.e., sub-LIRGs).
This universal trend illustrates the ``downsizing" scenario: infrared-bright massive galaxies gradually quench their extreme star formation, transitioning out of the SMG phase into passive evolution \citep{miller2015}, while lower-mass galaxies dominate star formation.

From another perspective, our sample at $z <$ 1 mainly consists of LIRGs, and at $z =$ 1-2, includes ULIRGs. However, their contribution is only comparable to $Herschel$-selected ULIRGs \citep{magnelli2013}, and significantly lower than the total SFRD. 
Several studies on the dust temperature and luminosity of SMGs provide useful insights: 850 $\micron$-selected SMGs tend to be cooler than $Herschel$-selected sources \citep{simpson2017, dud2020, liao2024} and exhibit lower average temperatures compared to local and low-$z$ samples \citep{dud2020}. At fixed luminosity, cooler SMGs show higher submillimeter fluxes \citep{swinbank2014, cunha2015}. 
This suggests that at lower redshifts ($z \lesssim$ 1–2), our current survey depth is likely to miss a population of 850 $\micron$–faint but warmer galaxies with lower dust masses (Section~\ref{sec:sed_bin}).  
Detecting such low-dust-mass systems at low redshift requires observations at shorter submillimeter wavelengths; for example, ultra-deep 450 $\micron$ surveys—such as STUDIES \citep{lim2020a}—are better suited to uncover this missing population.

In the redshift range $z =$ 2-2.5, the ASAGAO survey combined with archival data achieves a 1$\sigma$ depth of 0.1-0.2 mJy at 850 $\micron$ (assuming a dust emissivity index $\beta$=1.8), corresponding to $L_\text{IR} \sim 10^{11} L_\odot$ and resolving most of the cosmic infrared SFRD. The SSA22 deep field contributes about 25\% of this SFRD, approximately 1/3 to 1/4 of the ASAGAO results \citep{yamaguchi2020}.

At $z =$ 2.5-3.5, the SSA22 SMGs account for 50-60\% of the SFRD relative to \citet{madau2014}. For $z =$ 3-3.5, ALMA HUDF and S2CLS 850 $\micron$ archival data \citep{bourne2017, dunlop2017} reach a depth of $\sim$ 0.2 mJy (assuming $\beta$=1.8), revealing that faint submillimeter galaxies contribute significantly to cosmic star formation. 
Including the overdense region at $z =$ 3.09 (solid red dot in the figure), our sample contributes $\sim$ 62\% of the cosmic SFRD, consistent with these surveys.
However, if we exclude the contribution from the overdense regions (open circles, removing sources within the $3.09 \pm 0.05$ range), SMGs then account for approximately 25\% of the SFRD, consistent with the fraction observed at $z =$ 2–2.5. This indicates that overdense environments or large-scale structures significantly enhance the contribution to the infrared background.

As shown in the Figure~\ref{fig:fig4-6_sfrd}, the majority of the SMG sample reaches its peak contribution to the cosmic SFRD at $z \sim$ 2.5-3, approximately 1.5 Gyr earlier than the “cosmic noon”. Both ultra-deep surveys \citep{bourne2017, dunlop2017} and model \citep{zavala2021} suggest that dust-obscured star formation during this period accounts for about 80\% of the total SFRD.
\citet{dunlop2017} argue that the dominance of the dusty component in the cosmic SFRD during this epoch is primarily driven by the rapid increase in the number density of massive, dusty star-forming galaxies, with the evolution of galaxy dust fractions playing a secondary role.
This also suggests rapid evolution of ULIRGs in the early universe, where massive galaxies quickly accumulate stars and dust, eventually forming today's most massive elliptical galaxies and cluster members \citep{wang2019}.

\citet{dunlop2017, zavala2021, bouwens2022} further note that at $z >$ 4, dusty star-forming galaxies become increasingly rare, marking a transition in cosmic star formation activity from a dust-obscured to an unobscured mode. Their contribution to the cosmic SFRD declines sharply, falling to approximately 25–35\% at redshifts $z \sim 5$–6.
The SSA22 850 $\micron$-selected galaxies reflect this trend, with contributions declining rapidly to $\sim$ 20-25\% of the cosmic SFRD at $z >$ 3, consistent with other surveys.

For bright 850 $\micron$ sources (S$_\text{850}$ $>$ 3.9 mJy), we estimate their contribution to the SFRD and compare it with the AS2UDS sample (S$_\text{870}$ $>$ 3.6 mJy). As shown in the lower panel of Figure~\ref{fig:fig4-6_sfrd}, both samples yield consistent SFRD estimates. These bright SMGs still contribute $\sim$ 10-20\% of the cosmic SFRD at $z =$ 2-4.
%

At $z \gtrsim$ 4, ALMA and far-infrared detections become sparse.  Consequently, estimates of the dust-obscured SFRD at high redshifts often rely on a small number of bright sources and extrapolations of the infrared luminosity function \citep{madau2014, liu2018, grup2020}. The faint-end shape of the luminosity function in the early universe remains highly uncertain  \citep{traina2024}.
Studying dust-obscured star formation during this epoch requires a broader suite of observational approaches.

\citet{wang2019} identified a population of H-dropouts at $z \sim $ 4–6, which exhibit faint (sub)millimeter fluxes and are predominantly massive, dusty star-forming galaxies. Their contribution to the cosmic SFRD is comparable to that of brighter SMGs at the same epoch and amounts to approximately 10\% of the contribution from Lyman-break galaxies (LBGs).
However, when matched in stellar mass, the contribute of massive LBGs (shown as blue triangles in Figure~\ref{fig:fig4-6_sfrd}) 
is significantly lower than that of star-forming galaxies or SMGs, by $\sim$ 2 dex.
This indicates that in the early universe, star formation in the most massive galaxies was predominantly obscured by dust, and that dusty star-forming galaxies constitute the majority of the massive galaxy population \citep{wang2019}.
This result further demonstrates that at $z >$ 3, most of the most massive galaxies are indeed optically faint \citep{wang2019}.

Optically faint galaxies (OFGs) in ALMA-detected SMG samples typically reside at higher redshifts \citep{simpson2014, dud2020} and may be linked to the formation of compact, spheroidal structures in high-redshift massive galaxies \citep{smail2021}.
More recently, leveraging the high sensitivity of JWST, studies have analyzed a population of OFGs. These sources are often undetected by ALMA and include H-band–dropout or similarly extreme optical/near-infrared–faint systems \citep{williams2024}. 
The majority of OFGs are extended, high-redshift dusty star-forming systems \citep{wang2019, barrufet2025}, with SFRs range from 10 to 100 M$_{\odot}$ yr$^{-1}$ and stellar masses that are 1-2 dex lower than those of typical SMGs \citep{got2024, williams2024}.
OFGs were often entirely missed by HST and lie below the typical detection limit of ALMA (log$_{\rm 10}$($L_\text{IR}$/L$_{\odot}$) $<$ 12) \citep{williams2024}. 
%
At $z \lesssim$ 4-5, OFGs account for only a modest portion of the dust-obscured SFRD  ($\sim$ 26\%; brown circles in figure~\ref{fig:fig4-6_sfrd}). 
However, their contribution becomes substantial at $z \gtrsim$ 5-7, which comparable to previous estimates derived from extrapolated infrared luminosity functions, although ``little red dots" (LRDs) dominate the SFRD at $z >$ 6. 
The growing role of OFGs implies that past infrared/(sub)millimeter surveys may have missed up to half of the obscured star formation at these redshifts. High-redshift OFGs thus offer a promising avenue to place tighter constraints on cosmic SFRD in the post-reionization era \citep{williams2024}.


\begin{figure}
    \centering
    \begin{minipage}{\columnwidth}
    \includegraphics[width=\textwidth]{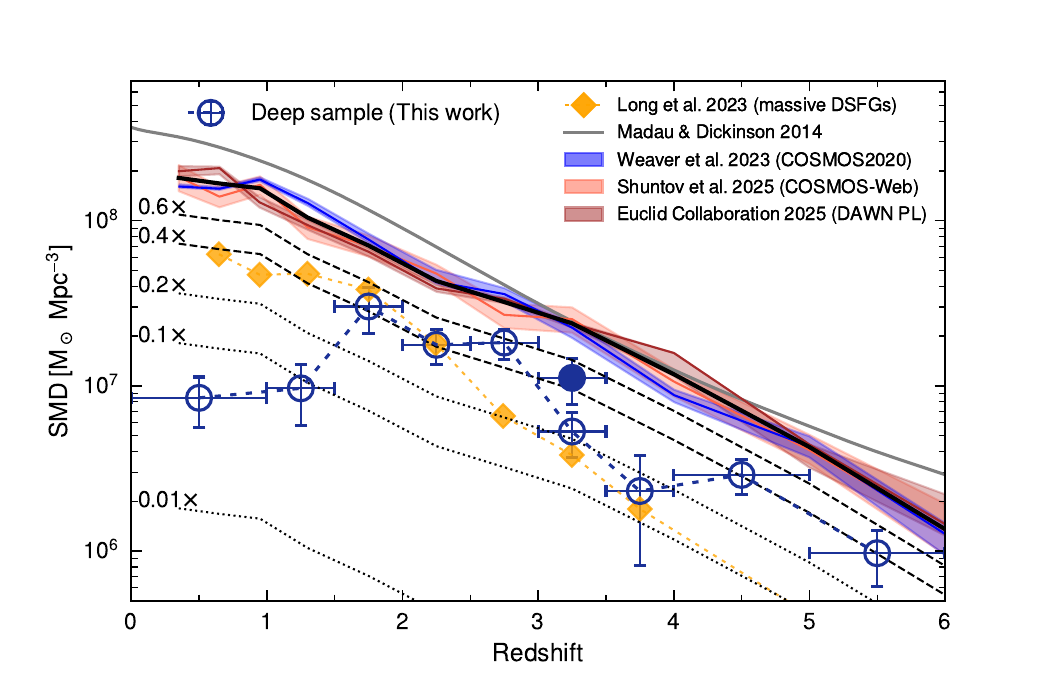}
    
    \caption{The contribution of SMGs to the cosmic stellar mass density. 
    We compare our results with the average measurements from three major survey programs \citep{euclid2025,shuntov2025,weaver2023}. The black solid, dashed, and dotted lines represent this averaged reference, with each line corresponding to a specific fractional contribution to the cosmic SFRD or SMD. 
    The contribution levels--ranging from 1\% to 60\%--are labeled on the left side of the respective lines.
    By averaging recent large survey data on cosmic SMD, we find that SSA22 SMGs contribute an average of $\sim$ 28\%. 
    Blue filled circles include contributions from SMGs within the redshift range of 3.09 ($\pm$ 0.05), again emphasizing significant contributions from overdense regions.}
    \label{fig:fig4-7_smd}
    \end{minipage}    
    
\end{figure}


\begin{figure}
    \centering
    \begin{minipage}{\columnwidth}
    \includegraphics[width=\textwidth]{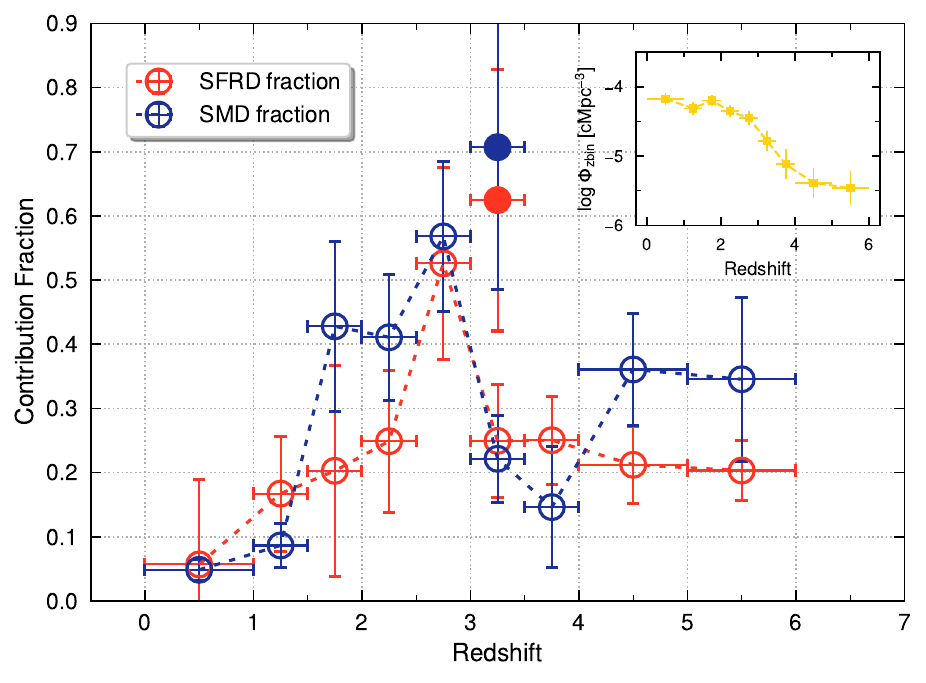}
    
    \caption{The fractional contribution of SMGs to the universe (solid symbols including overdense regions) as a function of redshift, peaking at 2.5–3 with values exceeding 50\%. 
    The inset in the upper right corner shows the number density of SMGs within the survey volume, which increases sixfold below redshift 4, reaching densities of about 2-3 $\times$ 10$^{-5}$ cMpc$^{-3}$ between redshifts 1-3. }
    \label{fig:fig4-8_contribution}
    \end{minipage}    
    
\end{figure}


\subsubsection{Contribution to the SMD}

Following the same methodology as described above, we calculate the cosmic SMD contribution of SMGs in the SSA22 deep field by summing the stellar masses of all SMGs within each redshift bin, with Poisson errors. 

Our results are compared with three major recent surveys (three colorful bands in Figure~\ref{fig:fig4-7_smd}): the $\sim$ 10 deg$^2$ Cosmic Dawn Survey Pre-launch by the Euclid telescope \citep{euclid2025}, the $\sim$ 0.5 deg$^2$ JWST COSMOS-Web survey \citep{shuntov2025}, and the $\sim$ 1 deg$^2$ COSMOS2020 project \citep{weaver2023}.
These surveys show discrepancies with \citet{madau2014}, who derived SMD by integrating the observed SFRD with a fixed conversion factor of 44\% (for a Chabrier IMF), accounting for stellar mass loss during star formation history. This approach may be affected by potential IMF evolution and baryon conversion efficiency, leading to differences between SFRD-derived and directly observed SMD. 
Furthermore, \citet{euclid2025} attribute the low-redshift discrepancies to improved data quality and uniformity, while the high-redshift differences may reflect a lag between rapid star formation and mass growth \citep{weaver2023}. However, it is more likely that these discrepancies are influenced by data and methodology: the high-redshift results of \citet{madau2014} were based on only two data points. Alternatively, systematic effects such as overestimated dust attenuation, uncertain UV-to-SFR conversion factors, and SED modeling uncertainties could play a role \citep{weaver2023, shuntov2025}.

On average, SSA22 SMGs contribute approximately 28\% to the cosmic SMD (Figure~\ref{fig:fig4-7_smd}).
At $z \sim$ 4, dust-obscured and unobscured star formation contribute comparably to the cosmic SFRD \citep{bourne2017, dunlop2017, zavala2021, bouwens2022}. 
\citet{zavala2021} suggested that at higher redshifts ($z =$ 4-6), the scarcity of dust-rich DSFGs results in a significant deficit of infrared emission relative to UV emission, indicating that unobscured star formation dominates cosmic star formation at these epochs. 
Despite their low number density at high redshift—approximately 3-4$\times$10$^{\rm -6}$ cMpc$^{\rm -3}$ at $z \sim$ 4–6—these rare, massive galaxies contribute $\sim$ 20–25\% to the cosmic SFRD and $\sim$ 35–40\% to the SMD (Figure~\ref{fig:fig4-8_contribution}).
  

At later epochs ($z <$ 4), dust-obscured star formation rapidly increases, dominating the cosmic SFRD and surpassing the contribution of UV-selected samples. Even among UV-bright galaxies with $M_{\rm star}$ $>$ 10$^{10}$ M$_{\odot}$, star formation is predominantly obscured \citep{bourne2017}. 

From the early universe to $z \sim$ 1–3, the comoving number density of SMGs increases by roughly an order of magnitude, reaching $\sim$ 4-6$\times$ 10$^{\rm -5}$ cMpc$^{\rm -3}$.
The SSA22 SMGs reach peak contributions at $z =$ 2.5–3.5, accounting for 50-60\% of the cosmic SFRD (including overdense regions) and a similar fraction of the SMD (Figure~\ref{fig:fig4-8_contribution}). 
\citet{long2023} used numerical models to estimate the contribution of massive DSFGs to the cosmic SMD, finding excellent agreement with observational SMG surveys over the redshift range $z \sim$ 2-4.

The number density of SMGs remains relatively stable between $z = 1.5$ and 2.5, yet the contribution of SMGs to the cosmic SFRD declines rapidly—becoming comparable to that at $z \sim 4–6$ (Figure~\ref{fig:fig4-8_contribution}).
This decline may be attributed to the overall slowdown in cosmic star formation and a reduced efficiency in converting baryons into stars.
During this epoch, the contribution of SMGs to the SMD ($\gtrsim$ 40\%) exceeds that to the SFRD. 
This discrepancy may arise from the substantial stellar mass accumulated by SMGs that are gradually quenching. Alternatively, the increased contribution from lower-mass galaxies may be delayed relative to the peak activity of SMGs, similar to the lag observed between cosmic noon and the peak of SMG activity. Uncertainties in the completeness of these large surveys could also contribute to this effect.

%
%
%

\citet{zavala2021} argued that the contribution of DSFGs to the cosmic SFRD remained relatively constant to the present day after reaching its peak ($\gtrsim 80\%$).
Moreover, \citet{long2023} suggest that the contribution of DSFGs to the SMD shows only mild evolution at $z \lesssim 2$.
However, as shown in Figure~\ref{fig:fig4-7_smd}, our SMG sample exhibits a rapid decline at $z < 1.5$, with the contribution of SMGs to both the SFRD and SMD dropping sharply to $\lesssim 10\%$. \citet{mck2025} report a sharp drop in the number density of SMGs. The contribution from bright SMGs also diminishes quickly after $z \sim 2.5$, becoming negligible by this epoch.
These observations collectively reflect the quenching of star formation in massive SMGs. Dust-obscured star formation—and the associated buildup of stellar mass—has progressively shifted toward lower-mass dusty systems (i.e., downsizing), whose individual contributions to the cosmic SFRD and SMD are small. 
Additionally, other populations of dusty galaxies not efficiently traced by 850 $\micron$ selection—such as intrinsically fainter, warmer systems with lower dust masses, or more compact systems—may dominate obscured star formation at late times but remain underrepresented in current submillimeter surveys.
(i.e., wavelength selection effects, section~\ref{sec:sfh_evolution})


\subsection{Evolution of SMGs}
\label{sec:sfh_evolution}

In the introduction, we provided a brief overview of the formation and evolutionary mechanisms of SMGs. 
In this section, we describe the evolution of SMGs from a phenomenological perspective, based on statistical observational results, and offer brief interpretive discussions.

Our analysis leads to the following synthesis:
In the early universe—around the epoch marking the end of reionization—massive DSFGs are exceedingly rare compared to UV-selected massive galaxies, and the infrared emission associated with dust-obscured star formation is far fainter than the cosmic ultraviolet/optical background.
At $z =$ 5-6, the comoving number density of dusty galaxies selected at 850\,$\micron$ and $\sim$1\,mm is approximately 3-4$\times10^{-6}$ cMpc$^{-3}$ \citep{mck2025} (Figure~\ref{fig:fig4-8_contribution}). The contribution from galaxies detected in typical 850\,$\micron$ surveys accounts for about 20\% of the total SFRD at these redshifts (Figure~\ref{fig:fig4-8_contribution}), which, relative to the results of \citet{madau2014}, represents an upper limit.
Moreover, \citet{zavala2021} argue that the total dust-obscured star formation activity still contributes $\sim$ 25-35\% to the SFRD at $z = 5$–6.

By $z \sim$ 4–5, star formation increasingly occurred within dense, dusty environments, surpassing unobscured modes in contribution \citep{dunlop2017}. This transition is reflected in the rising dust attenuation observed in both UV- and IR-selected galaxies (\citetalias{paperiii}), signaling a global shift from an unobscured to a dust-obscured regime \citep{bourne2017, dunlop2017, bouwens2020, bouwens2022, zavala2021}.
\citet{harikane2022a} suggest that galaxy star formation activity may be regulated by the assembly of dark matter halos.
During this epoch, a small subset of the most massive galaxies exhibited exceptionally high baryon conversion efficiencies \citep{xiao2024, merlin2025} (Figure~\ref{fig:fig4-1_spore}).

SMGs undergo intense star formation primarily at $z \sim$ 2–5 \citep{thomas2005}, likely fueled by gas-rich major mergers \citep{tacconi2008, engel2010, toft2014, chen2015, hodge2025}. Nevertheless, morphological and simulation studies indicate that the distribution of galaxy morphologies among SMGs resembles that of the general field population \citep{swinbank2010, casey2014, mcalpine2019, gillman2024}, suggesting their evolutionary pathways may not be fundamentally distinct. Indeed, a significant fraction ($\gtrsim$ 40–70\%) of SMGs appear to be driven by secular processes—such as smooth gas accretion and disk instabilities—rather than violent interactions \citep{magnelli2012, gillman2024, chan2025, zhang2025}. Observations consistently place SMGs at the high-mass end of the star-forming main sequence \citep{cunha2015, dunlop2017, lim2020a}.

The comoving number density of these galaxies increases rapidly below redshift $\sim$ 4 \citep{dunlop2017, fujimoto2024, traina2024}, rising by roughly an order of magnitude to $\sim$ 4-6$\times10^{-5}$ cMpc$^{-3}$ (Figure~\ref{fig:fig4-8_contribution}), when (sub)millimeter galaxies become abundantly detectable.
This is consistent with the conclusion of \citet{harikane2022a}, who attribute the rise in cosmic SFRD from $z \sim 10$ to $z \lesssim 4$ primarily to the rapid increase in the number density of dark matter halos.
By $z \sim$ 2-3, massive DSFGs reach their peak activity. The median redshift of 850\,$\micron$-selected SMGs lies in the range 2.3–2.7, and the mode of the redshift distribution in our sample is approximately 2.44. 
During this epoch, SMGs dominate cosmic star formation, contributing $\sim$ 50-60\% to the SFRD—a fraction that depends on survey depth.
Moreover, \citet{zavala2021} estimate that dust-obscured star formation as a whole accounts for $\sim$ 80\% of the total SFRD at this cosmic peak.

This peak activity is short-lived. Due to their extreme star formation rates, SMGs have gas depletion timescales of only tens to a few hundred Myr \citep{toft2014}. 
\citet{noeske2007a} argue that the global decline in the cosmic SFRD is primarily driven by the progressive exhaustion of gas reservoirs in galaxies.
In our sample, the modal gas consumption timescale is $\sim$ 200 Myr (Section~\ref{sec:lifetime}, Figure~\ref{fig:fig4-2_lifetime}). Environmental effects further accelerate quenching in overdense regions \citep{peng2010}.
Concurrently, below $z \sim$ 2–2.5, the baryon conversion efficiency within the massive dark matter halos hosting SMGs ($M_{\rm halo} \gtrsim 10^{13}\,h^{-1}\,M_\odot$) drops precipitously \citep{behroozi2013}. 
This decline is consistent with the “cold-mode” gas accretion model \citep{birnboim2003}: at $z \lesssim$ 2, inflowing gas in massive halos is typically shock-heated to the virial temperature \citep{dekel2006}, effectively cutting off the supply of cold gas and quenching further star formation.
Consequently, many massive SMGs ($M_* \sim 10^{11}\,M_\odot$) begin transitioning to quiescence during this epoch (Figure~\ref{fig:fig4-4_deltams}, Figure~\ref{fig:fig4-5_ssfr}), leading to a declining contribution to both the cosmic SFRD and stellar mass assembly.

At $z \sim 1.5$–2.5, although the number density of 850\,$\micron$ SMGs remains roughly constant, their contribution to both the cosmic SFRD and SMD declines rapidly, reaching approximately 20\% and 40\%, respectively—levels comparable to those at $z =$ 4-6 (Figure~\ref{fig:fig4-8_contribution}).
The results from \citet{harikane2022a} further indicate that from $z \sim 2$ to $z \sim 0$, cosmic expansion leads to a sharp decline in the halo accretion rate. Consequently, despite a modest increase in the star formation efficiency, the overall cosmic SFRD exhibits a declining trend.

In contrast, lower-mass galaxies ignite their main star formation episodes later \citet{noeske2007b}. At $z >$ 2, they exhibit lower metallicities \citep{cowie1996, nelan2005, thomas2005, pilyugin2011} and higher sSFR. 
Their star formation declines more gradually \citep{noeske2007b}, partly because stellar feedback in shallower potential wells prolongs gas retention and star formation \citep{neistein2006}.
Thus, by $z \lesssim$ 1.5–2.5, as massive galaxies transition from bursty to quiescent modes under the influence of “downsizing”, the locus of cosmic star formation shifts toward lower-mass systems \citep{cowie1996, brinchmann2000, juneau2005}, which sustain elevated sSFR over longer timescales \citep{noeske2007b}.

At $z \sim 1$–1.5, the contribution of SMGs declines rapidly, accompanied by a decrease in their number density \citep{mck2025}. By $z < 1$, 850\,$\micron$ and millimeter-wave surveys struggle to detect SMGs altogether.

In addition to the aforementioned mechanisms of “mass quenching”, “environmental quenching”, and “halo quenching”, AGN feedback also plays a non-negligible role in quenching star formation in massive galaxies.
Low-luminosity AGN are commonly detected in massive quiescent galaxies at $z \sim$ 2 \citep{olsen2013}, suggesting active involvement in quenching. \citet{merlin2025} report strong broad emission lines indicative of AGN activity in $\sim$1/3 of their massive quiescent sample. Stellar feedback, too, can act in concert: intense starbursts may heat and expel gas from even deep potential wells on Myr timescales \citep{merlin2012}.
Given that most observed SMGs are dominated by disk structures, morphological quenching (also known as Q-quenching) may occur if the growth of a central bulge stabilizes the disk, thereby suppressing gas fragmentation and star formation \citep{martig2009}.
Furthermore, rapid mass acquisition can also quench galaxy star formation. \citet{oser2010} showed that in cosmological simulations, stellar mass assembly appears to occur in “two-phase”: early $in situ$ star formation via cold accretion, followed by later structural growth through minor mergers that increase size and mass without reigniting star formation, thereby accelerating quenching.

\citet{brinchmann2000} note that fewer than 10\% of local massive ellipticals formed via recent ($z <$ 1) major mergers, implying most originated from high-redshift interacting systems. 
In the context of SMGs, one view holds that massive galaxies gradually evolve into compact quiescent systems at $z \sim$ 1–2 \citep{barro2013, dekel2014, chen2015}, and later into ellipticals via dry mergers \citep{toft2014}. %
However, observations complicate this picture. 
Recent studies reveal that the majority of SMGs are disk-dominated \citep{gillman2023, gillman2024, lebail2024, chan2025, hodge2025, zhang2025}. Concurrently, \citet{gillman2024} demonstrate that the stellar disk structures of SMGs can persist down to $z \sim$ 1. 
Theoretical work also indicates that disk rebuilding after a merger is feasible through mechanisms such as environmental gas accretion or other processes \citep{governato2009, barro2013}.
Do massive DSFGs primarily evolve into local elliptical galaxies through a sequence of compaction (e.g., major mergers, violent disk instability) followed by slow expansion via minor mergers? Or do a significant fraction evolve directly into massive late-type galaxies in the local universe? 
Alternatively, is the evolutionary landscape more complex, with multiple pathways coexisting rather than a single dominant channel? 
Resolving the evolutionary trajectories of massive galaxies is essential for understanding the present-day universe.

Finally, we discuss the “downsizing” phenomenon and the scarcity of SMGs detected at $z \lesssim$ 1 in submillimeter surveys.
A robust observational trend in galaxy evolution is “downsizing”: massive galaxies complete their star formation early (at high redshift), while lower-mass systems remain actively forming stars to lower redshifts. 
First articulated by \citet{cowie1996} to explain low metallicities in high-$z$ damped Lyman-$\alpha$ systems, this paradigm has been reinforced by numerous studies.
\citet{brinchmann2000} showed that most massive galaxies had already quenched by $z \sim$ 1, whereas low-mass galaxies continued forming stars. Environmental dependence was highlighted by \citet{thomas2005}, who found earlier quenching in denser regions. 
\citet{nelan2005} and \citet{pilyugin2011} observed a “downsizing” trend in metallicity as a function of galaxy mass.
Using a galactic hybrid evolutionary tracks model, \citet{firmani2010} successfully reproduced the “downsizing” signature reflected in the stellar mass–halo mass relation at various redshifts.
Overall, the downsizing phenomenon appears to be a general pattern in galaxy evolution, manifested in both early- and late-type galaxies, as well as in the red and blue sequences.
Why is downsizing so pervasive? It reflects two intertwined processes: (1) rapid, early star formation and quenching in massive systems, and (2) prolonged, lower-level activity in low-mass galaxies that dominates the SFRD at late times.
\citet{noeske2007b} framed this as “staged galaxy formation”: low-mass galaxies start forming stars later and exhibit longer e-folding times for SFR decline, resulting in higher sSFR at low redshift.
Massive galaxies, by contrast, experience early, intense bursts—potentially triggered by internal instabilities \citep{gillman2024} or enhanced merger rates in dense environments \citep{thomas2005}—and assemble their stars rapidly \citep{neistein2006}, exhausting their gas reservoirs early.

Observationally, SMGs become increasingly elusive at $z \lesssim$ 1. As shown in Section~\ref{sec:sfrd}, their contribution to the SFRD drops sharply below $z \sim$ 1.5, with LIRGs taking over, and sub-LIRGs dominating at even later times \citep{magnelli2013}. 
We attribute this decline to three primary factors:
First, quenching and gas exhaustion. In the massive halos ($M_{\rm halo} \gtrsim 10^{13}\,h^{-1}\,M_\odot$) typical of SMGs, baryon conversion efficiency plummets below $z \sim$ 2.5, with star formation efficiency (SFE) falling from $\sim$ 10\% to just a few percent \citep{behroozi2013, shuntov2022}. Coupled with short gas depletion times ($\lesssim$ few hundred Myr), most SMGs quench before $z \sim$ 1.5. Moreover, the cosmic molecular gas density peaks at $z \sim$ 1.5–2 and declines thereafter, with gas fractions dropping and depletion times increasing rapidly \citep{liu2019}.
This is reflected in the statistics: as redshift decreases, the maximum infrared luminosity of galaxies detected at 850 $\micron$ also declines.
Second, the downsizing effect itself: the most extreme, IR-luminous starbursts in massive galaxies fade, leaving lower-mass systems to dominate ongoing star formation \citep{miller2015}.
Finally, there is the issue of band selection effects.
The selection effect at 850 $\micron$ is tied to dust mass. Consequently, under finite detection limits, 850 $\micron$ surveys may miss sources that have higher dust temperatures and higher infrared luminosities but lower dust masses. Observations also indicate that local or low-redshift ULIRG samples generally exhibit higher dust temperatures than SMGs \citep{symeonidis2013, clements2018}.
Moreover, the negative $k$‑correction at 850 $\micron$ is weaker at lower redshifts; as redshift decreases, the observed flux density at 850 $\micron$ drops rapidly along the Rayleigh–Jeans tail. For a given infrared luminosity, SMGs with cooler dust temperatures show higher flux densities in the submillimeter bands \citep{swinbank2014, cunha2015}. Therefore, 850 $\micron$ observations are more sensitive to the colder dust component \citep{chapman2004, simpson2017, dud2020, liao2024}, a bias that becomes particularly pronounced at lower redshifts.
This implies that single-band 850 $\micron$ selection may suffer from incompleteness. Within the redshift range $z \lesssim$ 1-1.5, limited by current 850 $\micron$ detection depths, a substantial population of galaxies with relatively high dust temperatures but insufficient dust mass is likely missed.
This conclusion is consistent with multiple studies showing that longer-wavelength submillimeter/millimeter observations are more efficient at detecting high-redshift sources \citep{casey2014, zavala2014, wang2017, casey2021}. Shorter-wavelength far‑infrared/submillimeter bands predominantly probe low-redshift galaxies and contribute more strongly to the low-redshift end of the infrared luminosity function \citep{magnelli2013}.

\begin{figure}
    \centering
    \begin{minipage}{\columnwidth}
    \includegraphics[width=0.9\textwidth]{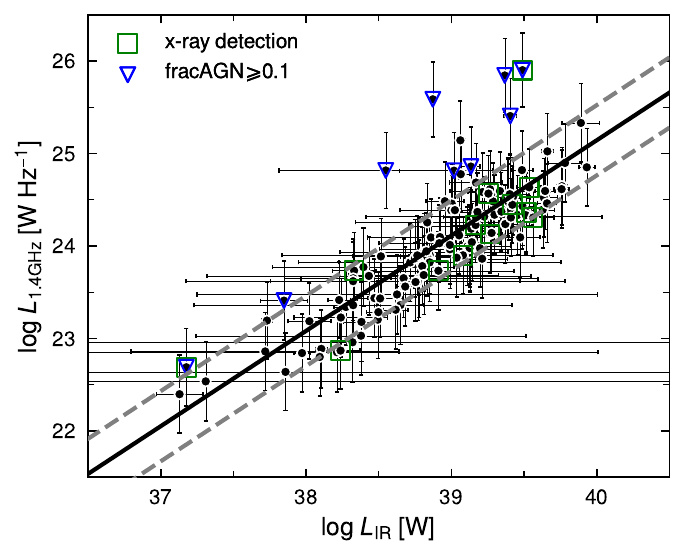}
    
    \caption{The tight correlation between radio luminosity and infrared luminosity of SMGs in the SSA22 field, both of which directly trace star formation activity within these galaxies. 
    The black solid line represents the linear regression fit, with gray dashed lines delineating the 1$\sigma$ scatter of 0.38 dex. 
    The open green squares represent sources detected in X-rays, while the open blue triangles indicate galaxies with an AGN fraction greater than 0.1 as inferred from SED fitting.
    %
    Galaxies with higher AGN fractions deviate due to excess radio emission. 
    Most X-ray detected SMGs remain within the correlation, showing no excess radio emission, suggesting that X-ray emissions in our SMG sample likely originate primarily from star formation. }
    
    \label{fig:fig5-1_ir_radio}
    \end{minipage}
    
\end{figure}

\section{IR-Radio Correlation}
\label{sec:irrc}

In this section, we present the relationship between radio and submillimeter emission for our SMG sample and characterize the distribution of the quantitative parameter $q_{\rm IR}$ (Section~\ref{sec:irrc_q}). 
The IRRC is remarkably robust; as briefly reviewed in Section~\ref{sec:intro}, it appears to hold out to high redshifts. 
However, whether the IRRC—or equivalently, the $q_{\rm IR}$ parameter—undergoes cosmic evolution remains debated. We address this question using our data and offer a brief discussion (Section~\ref{sec:irrc_does_evolve}). Finally, we examine the ratio of submillimeter to radio flux densities (Section~\ref{sec:irrc_flux_ratio}).

We must also caution about potential misidentifications in the sample, which are discussed in detail in \citetalias{paperiii}. Sources with redshifts below $z \sim 1$ may arise from spurious cross-matching. 
Moreover, since our sample identification relies heavily on radio counterparts, this selection could introduce additional systematic effects in the present analysis \citep{algera2020a}. For instance, it is possible that some radio-bright, mid-infrared luminous but submillimeter-faint warm galaxies have been inadvertently included in our sample \citep{chapman2004}.
Additionally, it should be noted that we assumed a fixed radio spectral index of $\alpha = -0.8$ when converting 3 GHz flux densities to 1.4 GHz. While studies of large-sample SFGs suggest a median spectral index around $-0.8$ with no strong redshift evolution \citep{an2021}, adopting a single value may introduce systematic errors. For instance, spectral aging could lead to an underestimation of high-frequency flux densities when extrapolating from lower frequencies \citep{thomson2019}.



\begin{figure*}
    \centering
    \includegraphics[width=0.9\textwidth]{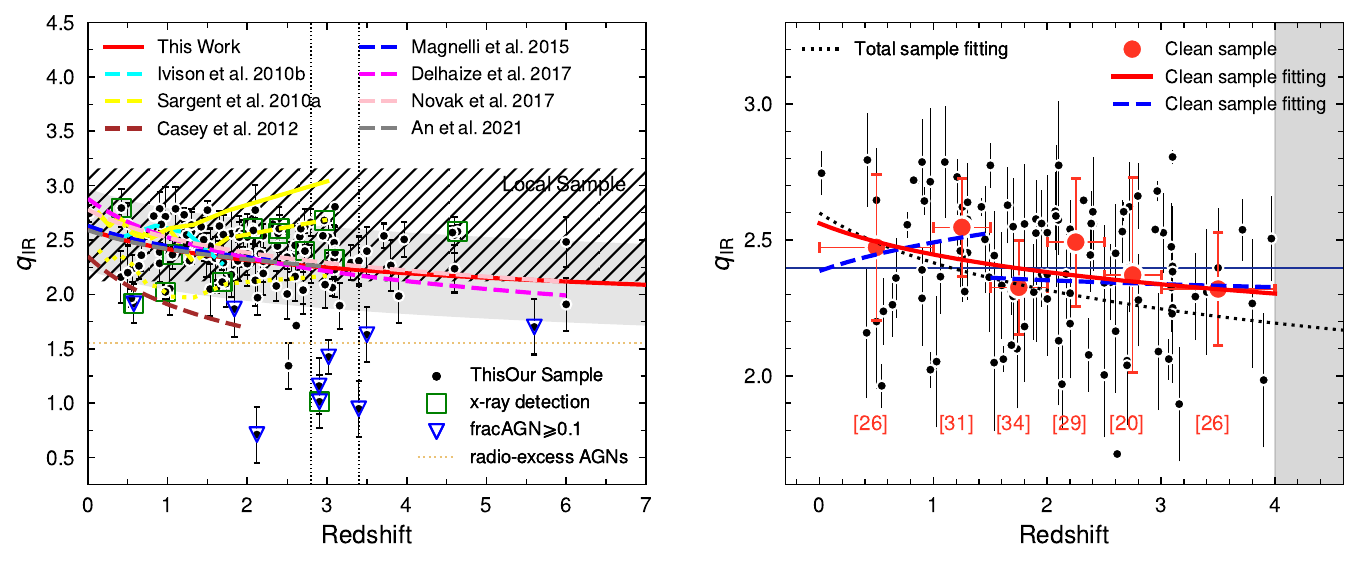}
    
    \caption{Left: Evolution of the infrared-radio correlation parameter $q_\text{IR}$ with redshift for the SMG sample.The parameter decreases with redshift as $q_\text{IR} \propto (1+z)^{-0.11}$, with a median value of 2.37. The gray shaded region indicates the 1$\sigma$ uncertainty of the fit, while the hatched band shows the mean and 2$\sigma$ range for local galaxy sample. Vertical dotted lines delineate the range $z = 3.09 \pm 0.3$, and horizontal dotted line below define as radio-excess AGNs  ($q_\text{IR} \leqslant$ 1.55; \citealp{algera2020b}), suggesting that overdense regions may promote galaxy evolution.
    Right: Evolution of $q_\text{IR}$ with redshift for the cleaned sample (removing sources with AGN fractions $\geqslant$ 0.1 and $z >$ 4). 
    Red circles indicate the average $q_\text{IR}$ in different redshift bins, with the number of sources in each bin shown in brackets. The red solid line shows the fit for the cleaned sample, which is shallower than the full-sample fit (black dotted line). The blue dashed line represents the fit for the cleaned sample split into $z <$ 1.5 and $z$ = 1.5-4, revealing no strong evolution within these intervals. The overall trend may arise from intrinsic differences between high- and low-redshift SMGs, possibly influenced by downsizing.
    The horizontal solid line represents the median $q_\text{IR}$ of cleaned sample. }
    
    \label{fig:fig5-2_qir}
\end{figure*}


\subsection{The IR-Radio Correlation Factor $q_\text{IR}$}
\label{sec:irrc_q}

Because both far-infrared and radio emission originate from massive stars, galaxies exhibit a remarkably tight linear correlation between their total infrared and 1.4 GHz radio luminosities over several orders of magnitude \citep{helou1985}—a relation known as the infrared–radio correlation.

\citet{helou1985} introduced the parameter $q$ (or $q_\text{FIR}$) to quantify the IRRC, defined as the logarithmic ratio of far-infrared flux (42.5-122.5 $\micron$) to 1.4 GHz radio flux \citep{condon1992, yun2001}. However, with evolving observational practices, the correlation is now commonly expressed using the total infrared luminosity ($L_\text{IR}$, 8–1000 $\micron$) and the 1.4 GHz radio luminosity (or flux), denoted as $q_\text{IR}$ (or $q_\text{TIR}$, total infrared luminosity). On average, $q_\text{IR}$ = $q_\text{FIR}$ + log$_\text{10}$(1.91) \citep{magnelli2015}.

We first calculate the rest-frame 1.4 GHz luminosity, $L_\text{1.4 GHz}$. 
Using the relation $\nu_{\text{rest}}/\nu_{\text{obs}} = (1+z)$ and $S_{\nu} \propto \nu^{\alpha}$ (radio spectral index $\alpha$, assumed to be -$0.8$), we derive the observed 1.4 GHz flux $S^{\text{obs}}_{\text{1.4 GHz}}$ from the observed 3 GHz flux. 
Following \citet{ivison2010b}, we apply a $k$-correction factor $K$ to convert $S^{\text{obs}}_{\text{1.4 GHz}}$ to the rest-frame flux: $K \times S^{\text{obs}}_{\text{1.4 GHz}}$. Since the observed 1.4 GHz corresponds to a rest-frame frequency of 1.4(1+$z$) GHz, the $k$-correction is $K = S^{\text{rest}}_{\nu}/S^{\text{obs}}_{\nu} = \nu^{\alpha}/[\nu(1+z)]^{\alpha} =  (1+z)^{-\alpha}$.
The rest-frame 1.4 GHz luminosity is then:
\begin{equation}
    L_{1.4\,\mathrm{GHz}} = \frac{4 \pi D_L^2}{(1+z)^{1+\alpha}} \, S^{\mathrm{obs}}_{1.4\,\mathrm{GHz}},
\end{equation}
where $D_L$ is the luminosity distance in meters. The additional $(1+z)^{-1}$ term accounts for the frequency element interval compression due to cosmological redshift \citep{murphy2009a, ivison2010b}.

Figure~\ref{fig:fig5-1_ir_radio} demonstrates the tight relationship between radio (1.4 GHz) and infrared (8-1000 $\micron$) luminosities.
A linear regression fit yields: $L_{\text{1.4 GHz}} \, [\text{W} \, \text{Hz}^{-1}] = (1.03 \pm 0.05) \times L_{\text{IR}} \, [\text{W}]+ (-16.11 \pm 2.00)$, with a 1$\sigma$ scatter of 0.38 dex. 
This result is consistent with that derived from local galaxy samples \citep{bell2003}, indicating that high-redshift SMGs follow the IRRC and lie at the high-luminosity end. A few outliers are present, most of which are likely due to excess radio emission from AGN activity.

We adopt the $q_\text{IR}$ definition from \citet{ivison2010b} and \citet{sargent2010b}:
\begin{equation}
    q_{\text{IR}} = \log_{10} \left( \frac{L_{\text{IR}} \, [\text{W}]}{3.75 \times 10^{12} \, \text{Hz}} \right) - \log_{10} \left( L_{\text{1.4 GHz}} \, [\text{W} \, \text{Hz}^{-1}] \right),
\end{equation}
where $L_\text{IR}$ is the rest-frame integrated infrared luminosity (8–1000 $\micron$), and $L_\text{1.4 GHz}$ is the rest-frame 1.4 GHz luminosity. The factor $3.75 \times 10^{12}$ Hz corresponds to the midpoint frequency of the 42.5–122.5 $\micron$ band \citep{ivison2010b}.

In Figure~\ref{fig:fig5-2_qir}, our analysis of the 146 sources jointly detected in the submillimeter and radio bands yields a median IRRC coefficient of $q_\text{IR}$ = 2.37 $\pm$ 0.04 with a 1$\sigma$ dispersion of 0.37. 
The sample exhibits a mild decreasing trend with redshift: $q_{\text{IR}} = (2.60 \pm 0.11) \times (1+z)^{-0.11 \pm 0.04}$. 

Blue inverted triangles and green squares in Figure~\ref{fig:fig5-2_qir} denote sources with an AGN fraction of at least 10\% and those detected in X-rays, respectively. The AGN-tagged sources are predominantly outliers, exhibiting excess radio emission (Figure~\ref{fig:fig5-1_ir_radio}) and consequently lower $q_\text{IR}$ values. 
These sources fall below or near the radio-excess AGNs threshold (yellow dashed line in Figure~\ref{fig:fig5-2_qir}, $q_\text{IR} \leqslant$ 1.55; \citealp{algera2020b}), while several unmarked sources close to this criterion may also be potential AGN candidates.
Interestingly, the sources exhibiting the lowest $q_\text{IR}$ values are located around $z \sim 3.09$, potentially associated with the SSA22 protocluster (if consider photometric redshift uncertainties $\Delta z \sim 0.3$). This suggests that dense environments may enhance galaxy interactions and/or leverage the abundant gas reservoirs within clusters promote galaxy evolution, thereby boosting AGN activity \citep{lehmer2009, umehata2015}.
Among the AGN-flagged sources, only two have X-ray detections. Most X-ray-detected sources instead display near- or above-average $q_\text{IR}$, exhibiting elevated infrared luminosities. This implies that X-ray emission in DSFGs predominantly originates from vigorous star formation (e.g., X-ray binaries) rather than AGN activity.

Compared to local galaxy samples ($q_\text{IR}$ $\sim$ 2.64; \citealp{yun2001, bell2003}), our SMGs show systematically lower values by 0.27 dex, implying either reduced infrared luminosities or enhanced radio emission by a factor of $\sim$ 2. Nevertheless, the bulk of our SMG population remains within 2$\sigma$ of the local distribution (shaded region in left panel of Figure~\ref{fig:fig5-2_qir}).
These results align well with previous studies of infrared/radio-selected galaxies \citep{murphy2009a,ivison2010a,ivison2010b,sargent2010a,magnelli2015, delhaize2017} and SMG samples \citep{murphy2009a}, which report median $q_\text{IR}$ values in the range $\sim$ 2.34-2.41 and redshift evolution slopes between -0.10 and -0.26. Our findings show particularly close agreement with the large infrared/radio-selected sample of \citet{magnelli2015,an2021}, sharing similar distributions and evolutionary trends (slope $\sim$ -0.10 to -0.12) at 2.7-3$\sigma$ significance \citep{magnelli2015}.

The lower overall $q_\text{IR}$ distribution and steeper evolution reported by \citet{casey2012}, as they note, may stem from selection biases in their radio-detected sample, which is corroborated by result from radio-detected subsample of \citet{sargent2010a}  (yellow dotted line in left panel of Figure~\ref{fig:fig5-2_qir}). 
\citet{sargent2010a} present $q_\text{IR}$ distributions for samples selected solely in the radio (yellow dotted line), in the infrared (solid line), and via combined detection (dashed line). At $z <$ 1.5, the infrared- and combined-selected samples show similar trends, while the radio-selected sample already deviates by $\sim$ 0.3 dex. Notably, the radio-selected results align closely with those of \citet{casey2012}. This confirms that $q_\text{IR}$ derived from radio-selected samples alone can be significantly biased—systematically lower and exhibiting a steeper evolutionary slope \citep{delhaize2017}.
At higher redshifts, the discrepancy between the radio-selected and combined-selected samples grows to 0.5 dex or more, while the infrared-selected sample also diverges significantly from the combined sample by $\sim$ 0.3 dex. 
At same time, the $q_\text{IR}$ distribution of their combined-detection sample diverges from our results and those in the literature at $z >$ 1.5, and the scatter in the data increases with redshift (Table 6 in \citealt{sargent2010a}), likely due to the lack of reliable photometric redshifts in this regime \citep{sargent2010a, delhaize2017}.

We re-analyzed a sample from the SSA22 deep field, excluding sources likely affected by radio-loud AGN (i.e., those with AGN fraction $\geqslant$ 0.1, mostly outliers) and high-redshift sources ($z >$ 4) where detection completeness may be an issue. 
For this ``cleaned sample", the median $q_\text{IR}$ is 2.40 $\pm$ 0.04 . A fit to this sample yields a redshift evolution slope $\gamma$ = -0.07 $\pm$ 0.03 (where $q_\text{IR}$ $\propto$ (1+z)$^\gamma$), with a significance of only $\sim$ 2.2$\sigma$ (red solid line in right panel of Figure~\ref{fig:fig5-2_qir}). This indicates that, after removing AGN-dominated outliers, the overall evolutionary trend becomes flatter and less significant.

Our results are also consistent with those for SMG samples from \citet{murphy2009a}.
\citet{michalowski2010} noted a significant difference in the median $q_\text{IR}$ between sources at $z >$ 1.5 and $z <$ 1.5, despite claiming no strong evolution in SMGs overall.
Motivated by this, we derived median $q_\text{IR}$ values of this cleaned sample in each redshift bin (shown as red circles in right panel of Figure~\ref{fig:fig5-2_qir}). 
For sources at $z <$ 1.5, the median $q_\text{IR}$ is 2.55 $\pm$ 0.07, close to the local value \citep{bell2003}. For the intermediate redshift bin $z =$ 1.5-4, the median $q_\text{IR}$ is 2.36 $\pm$ 0.04 values are in close agreement with \citet{michalowski2010} (see their Figure 6). Fitting the $z <$ 1.5 and $z =$ 1.5-4 subsamples separately yields slopes of 0.06 $\pm$ 0.08 and -0.02 $\pm$ 0.07, respectively (blue dashed lines in right panel of Figure~\ref{fig:fig5-2_qir}), suggesting no clear evolution within either bin.

We therefore conclude that the overall evolutionary trend in $q_\text{IR}$ primarily arises from intrinsic differences between high-redshift SMGs and their lower-redshift ($z <$ 1.5) counterparts. Interestingly, $z \sim$ 1.5 corresponds to a critical epoch discussed earlier (Sections~\ref{sec:sfh_ms} and~\ref{sec:sfh_evolution}): the period when 850 \micron-selected SMGs begin quenching and downsizing shifts the population toward lower-mass systems. 
The underlying cause of difference may be that low-redshift SMGs possess intrinsic properties more similar to local galaxies and possibly exhibit additional infrared emission from cirrus components \citep{yun2001, michalowski2010}, whereas high-redshift systems display distinct characteristics shaped by their extreme star-formation environments.
This reinforces the idea that SMGs, selected at a specific wavelength, are not a homogeneous population but encompass a diverse range of galaxies \citep{grup2019}; this diversity will be further explored in Section~\ref{sec:irrc_flux_ratio}.

For radio-undetected sources via stacking analysis, \citet{thomson2014} derived a median $q_\text{IR}$ of $\sim 2.35$ for the ALESS sample. After further restricting the sample to $z \gtrsim 1.5$ and removing radio-excess SMGs, the median $q_\text{IR}$ remains at 2.33, which is remarkably close to our results.
In contrast, the large-sample study of SMGs from the AS2UDS survey reported a lower median $q_\text{IR}$ of $\sim 2.20$, which shows no evolution with redshift or other physical properties \citep{algera2020a}. The authors argue that SMGs exhibit a different IRRC compared to local galaxies, likely due to a combination of strong magnetic fields, high interstellar medium densities, and additional radio emission from secondary cosmic rays. Local ULIRGs share comparable magnetic field strengths with SMGs but reside in even denser environments.
\citet{thomson2019} pointed out that spectral aging, manifesting as convex spectral behavior, is present in some SMGs. This effect may result from young central starbursts and suppressed radio emission due to weaker magnetic fields, and could serve as evidence for $q_\text{IR}$ evolution. Indeed, \citet{thomson2014} noted that $q_\text{IR}$ co-evolves with the radio spectral index $\alpha$, offering a potential tracer of the evolutionary stage of high-redshift starburst galaxies.
A lower obscured fraction of star formation at early cosmic times may also contribute to the evolution of $q_\text{IR}$. As highlighted by \citet{shen2022}, while radio emission can respond on short timescales, dust buildup and stellar population evolution operate over significantly longer timescales.

\begin{figure*}
    \centering
    \includegraphics[width=0.9\textwidth]{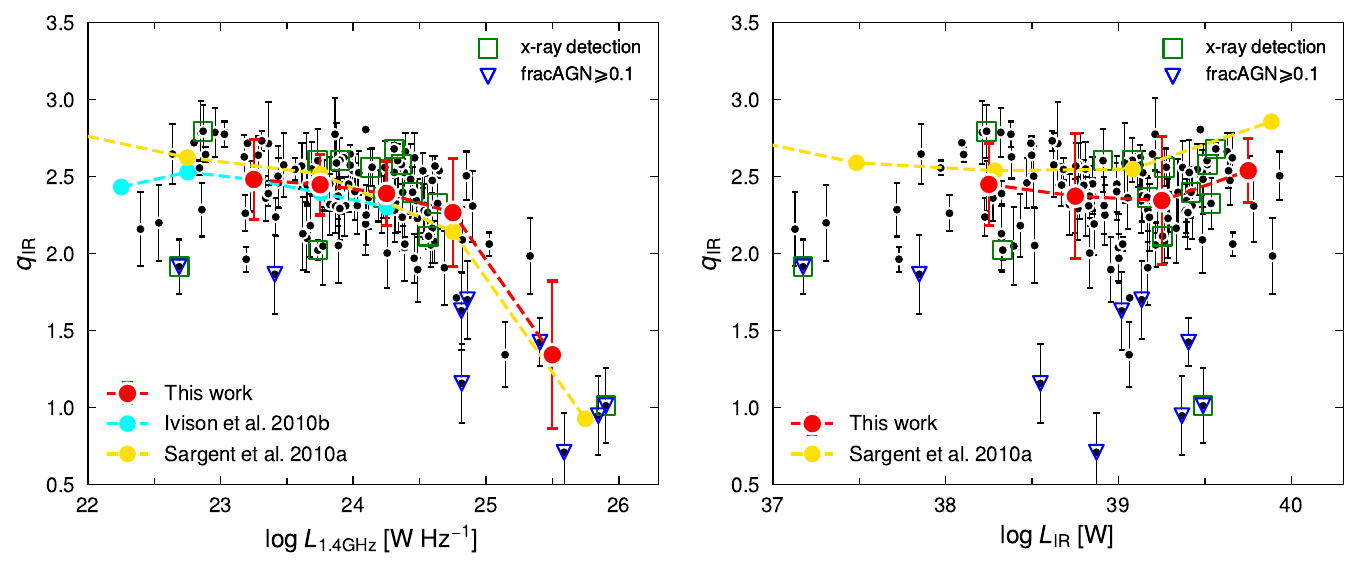}
    \caption{Left: Variation of the $q_\text{IR}$ with radio luminosity.
    A weak declining trend is observed, with a rapidly declines above 10$^{24.5}$ W Hz$^{-1}$, suggesting  that the most radio-luminous systems are dominated by extremely bright AGNs. 
    Right: Variation of $q_\text{IR}$ with infrared luminosity.
    At the brightest luminosity range ($L_{\text{IR}} \gtrsim 10^{39.5} \, \text{W}$), $q_\text{IR}$ increases, possibly indicating additional processes enhancing infrared emission or a relative lag in radio output in extreme starburst environments. }
    
    \label{fig:fig5-3_qir_l1p4lir}
\end{figure*}

We compare $q_\text{IR}$ with both radio and infrared luminosities (Figure~\ref{fig:fig5-3_qir_l1p4lir}).
Our results show that the trend of $q_\text{IR}$ with 1.4 GHz radio luminosity ($L_\text{1.4 GHz}$) in SMGs is in good agreement with previous studies \citep{ivison2010b, sargent2010a}, and this trend is also observed in nearby galaxies selected in the optical, infrared, and radio bands \citep{moric2010}. A slight decline in $q_\text{IR}$ is seen at $L_\text{1.4 GHz} \lesssim 10^{24.5}$ [W Hz$^{-1}$], which becomes more pronounced at higher luminosities and drops sharply for sources exceeding $10^{25}$ [W Hz$^{-1}$]. Most sources above this radio luminosity threshold are flagged as AGN, suggesting that such high radio luminosities are likely dominated by powerful AGNs, which typically exhibit low $q_\text{IR}$ values.
In the highest infrared luminosity bin ($L_\text{IR} \sim 10^{39.5}$ [W]), $q_\text{IR}$ shows a tendency to increase with $L_\text{IR}$, consistent with \citet{sargent2010a}. This may indicate the presence of additional infrared emission mechanisms in galaxies with the most extreme star formation activity, or a relatively slower increase in radio emission compared to infrared.
At $L_\text{IR} \sim 10^{38.2}$ [W], $q_\text{IR}$ appears slightly higher than at intermediate luminosities, these sources are predominantly at $z \lesssim 1.5$, potentially reflecting the additional contribution from cirrus emission in low-redshift galaxies \citep{yun2001}. However, this deviation is not statistically significant, and we conclude that $q_\text{IR}$ is largely independent of infrared luminosity.


\subsubsection{Does the Infrared-Radio Correlation Evolve?}
\label{sec:irrc_does_evolve}

The question of whether $q_\text{IR}$ truly evolves with redshift remains actively debated.

%
\citet{sargent2010a, sargent2010b} argue that apparent evolutionary trends stem from selection biases rather than intrinsic physical changes. The detection limitation of infrared or radio surveys can introduce biases in high-redshift samples, affecting the observed $q_\text{IR}$ distribution. For instance, sensitivity limits in the infrared may bias high-redshift samples toward extreme starbursts or AGN-dominated systems, while radio detection limits can increase the fraction of radio-loud AGNs in the sample \citep{ivison2010b}.
Additionally, \citet{sargent2010a} contend that the failure to properly account for confusion from undetected faint sources below the detection threshold has led to the conclusion of IRRC evolution. They assert that the IRRC is essentially invariant out to $z \lesssim 1.4$, although \citet{magnelli2015} note this conclusion may still be limited by the small number of high-redshift sources available.
Some studies report a correlation between $q_\text{IR}$ and stellar mass, with $q_\text{IR}$ decreasing as mass increases \citep{delvecchio2021, an2024}. Also,  \citep{bell2003} suggest that the ratio of thermal to non-thermal radio emission may depend on galaxy mass. \citet{delvecchio2021} propose a more radical view, suggesting that the apparent redshift evolution of $q_\text{IR}$ is primarily driven by an incompleteness bias against low-mass galaxies at high redshift. They argue that the IRRC dependence on redshift is weak and that evolution is primarily driven by stellar mass.
However, our sample shows no evidence for a dependence of $q_\text{IR}$ on stellar mass (slope $\theta = -0.05 \pm 0.13$, assuming $q_{\text{IR}} \propto (\log M_{*})^{\theta}$), likely because our sample is dominated by massive systems ($\log(M_{*}/M_\odot) \sim$ 10.6–12).

Nonetheless, substantial evidence supports genuine $q_\text{IR}$ evolution, potentially driven by a combination of the following factors.

The most prominent factor may be AGN activity.
Studies indicate a substantial AGN fraction within SMG samples \citep{alexander2005, laird2010, johnson2013}. \citet{delhaize2017} find that AGNs exhibit significantly lower $q_\text{IR}$ values than star-forming galaxies (SFGs), with more luminous AGNs showing even lower $q_\text{IR}$ and steeper evolutionary trends.
By performing morphological classification of optically selected galaxies at relatively low redshifts ($z \lesssim 1.5$), \citet{molnr2017} find that spheroid-dominated systems exhibit significant evolution, and they are more likely to be contaminated by AGN.
As described above, AGN-tagged SMGs typically show excess radio emission and correspondingly lower $q_\text{IR}$.
Moreover, the majority of SSA22 SMGs have 1.4 GHz luminosities exceeding $10^{22.7}$ [W Hz$^{-1}$], a regime where the radio luminosity function is increasingly dominated by AGNs \citep{ivison2010b}.
However, does AGN activity always lead to lower $q_\text{IR}$? \citet{murphy2009a, moric2010, sargent2010a} note that most AGNs actually follow the same IRRC as star-forming galaxies, with $q_\text{IR}$ values similar to those in the local universe. This implies that $q_\text{IR}$ is ineffective at distinguishing obscured or radio-quiet AGNs \citep{murphy2009a}.

\citet{ivison2010a, ivison2010b} suggest that radio activity in high-redshift galaxies may be enhanced relative to infrared emission.
The efficiency of free-free (bremsstrahlung) cooling for cosmic-ray electrons may increase at high redshift, thereby boosting the contribution of thermal radio emission \citep{bell2003}, leading to a decrease in $q_\text{IR}$ with redshift. However, \citet{delhaize2017} argue that this mechanism alone cannot fully explain the observed trend (under standard assumptions, free-free emission contributes $\sim$ 10\% of the radio flux \citep{condon1992}).
\citet{bell2003} note that non-thermal radio emission dominates at low frequencies ($\lesssim$ 5 GHz), while thermal radio emission becomes more significant at higher frequencies ($\gtrsim$ 5 GHz). As redshift increases, a fixed observed frequency probes higher rest-frame frequencies. At the typical peak redshift of SMGs ($z \sim 2.6$), the observed 1.4 GHz already corresponds to rest-frame frequencies where thermal emission contributes more substantially. This enhanced free-free emission increases the total radio luminosity, resulting in a lower $q_\text{IR}$.

Physical differences between high-redshift SMGs and local galaxies may lead to different $q_\text{IR}$ values.
High-redshift SMGs are not direct analogs of local ULIRGs \citep{menendez2009}. SMGs encompass a diverse population with a wide range of physical properties and a large intrinsic scatter in $q_\text{IR}$.
Moreover, SMGs host extreme environments and physical processes.
For example, \citet{murphy2009a} describe gas ``bridges" forming between interacting SMG pairs (interactions and mergers are common in SMGs), which contain significant cosmic-ray electrons and magnetic fields, producing bright radio continuum emission. This radio emission can account for up to half of the system's total radio output, even before substantial star formation is triggered in the bridge. Such systems exhibit excess radio emission, leading to low $q_\text{IR}$ values.
Additionally, extreme starburst systems in SMGs may possess mechanisms that enhance synchrotron efficiency. \citet{michalowski2010} suggest that SMGs may be ``puffy starbursts"  (vertically and radially extended, with scale heights of $\sim$1 kpc), where reduced bremsstrahlung and ionization losses allow cosmic-ray electrons to produce stronger radio emission.
\citet{thomson2019} suggested that a young central starburst combined with a weak magnetic field can suppress radio emission, leading to spectral aging. Meanwhile, \citet{thomson2014} proposed that the infrared-to-radio ratio $q_\text{IR}$ co-evolves with the radio spectral index alpha. Additionally, an initially low obscuration fraction during early star formation may also contribute to the observed evolution of $q_\text{IR}$. As citet{shen2022} pointed out, radio emission can respond on short timescales, whereas dust buildup and stellar population maturation require significantly longer timescales.

Sample selection biases are also a critical factor that can result in an evolution trend of $q_{\rm IR}$.
Detection limitation mean that radio-selected samples miss radio-faint sources \citep{ivison2010b}, implying that such samples may preferentially include sources with stronger radio emission, leading to systematically lower $q_\text{IR}$ values. Similarly, high-redshift infrared-selected samples are biased toward detecting the most luminous infrared galaxies (e.g., ULIRGs) \citep{sargent2010a}, which may host more extreme star formation or AGN activity. 
As previously noted, \citet{sargent2010a} demonstrate significant differences in $q_\text{IR}$ between samples selected solely by radio, solely by infrared, and by combined detection, with these discrepancies growing at higher redshifts. The results of \citet{casey2012} support this interpretation.
Differences in $q_\text{IR}$ may also stem from unaccounted mass selection biases. \citet{delvecchio2021} argue that radio emission exhibits a stronger dependence on stellar mass than infrared emission, with more massive galaxies showing lower $q_\text{IR}$. Since SMGs are inherently biased toward massive systems, this could naturally explain differences in $q_\text{IR}$ compared to local samples.
The physical basis for a mass dependence could be that more massive galaxies confine cosmic-ray electrons more effectively, leading to higher non-thermal radio emission relative to their infrared output compared to lower-mass galaxies \citep{bell2003}. It may also be related to star formation rate surface density \citep{delvecchio2021}.

Finally, studies suggest that IRRC evolution may also correlate with galactic magnetic field strength, gas density, and star formation rate surface density \citep{magnelli2015, delvecchio2021, yoon2024}. Beyond intrinsic physical causes, differences in assumptions and methodologies can also introduce variations in $q_\text{IR}$: for example, \citet{delhaize2017} note that star-forming galaxies may have complex SEDs. Furthermore, assumptions about a fixed radio spectral index, dust emissivity index, and the impact or bias from $k$-corrections can all influence the derived $q_\text{IR}$ values \citep{murphy2009a}.

\begin{figure}
    \centering
    \begin{minipage}{\columnwidth}
    \includegraphics[width=\textwidth]{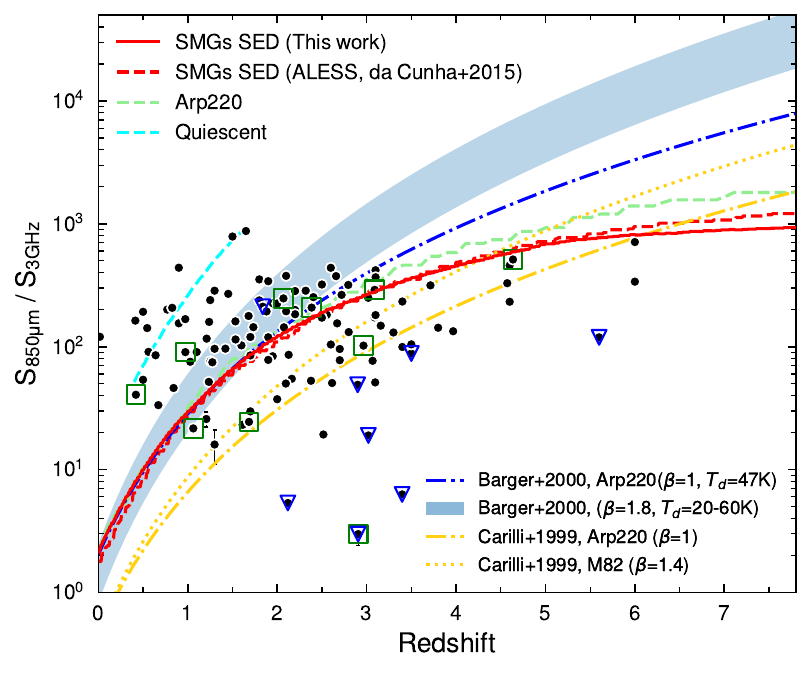}
    
    \caption{Ratio of submillimeter-to-radio flux as a function of redshift. 
    At low redshifts, SSA22 SMGs show a modest resemblance to templates of quiescent galaxies, though the fit is not particularly tight, while at higher redshifts they approach the ratios of warmer systems like Arp220. 
    Due to the intrinsic diversity of SMGs and the sensitivity of flux ratios to various physical conditions, neither a single template nor the average SMG SED adequately constrains the observed distribution, highlighting limitations in applying generalized models to individual sources.}
    \label{fig:fig5-4_flux_ratio}
    \end{minipage}  
    
\end{figure}

\subsection{Submm/Radio Flux Ratio}
\label{sec:irrc_flux_ratio}

Due to the historical lack of reliable redshift measurements for high-redshift galaxies, several studies have explored the relationship between submillimeter-to-radio flux ratios and redshift as a method for estimating redshifts, known as ``millimetric redshifts" \citep{cy99, barger2000, barger2012, yun2012}.
For instance, \citet{cy99} employed a simple double-power-law model, assuming a constant spectral index in the radio regime and approximating the submillimeter spectrum with an exponential form (Rayleigh-Jeans approximation), adopting submillimeter spectral indices $\alpha_s$ (= 2 + $\beta$, where $\beta$ is the dust emissivity index) of 3.4 and 3.0 for M82 and Arp220, respectively. \citet{barger2000} constructed a modified blackbody model based on Arp220 ($\beta$ = 1, $T_d$ = 47K). The predictions from these templates are shown in Figure~\ref{fig:fig5-4_flux_ratio}.

Additionally, we use templates representing high-redshift SMGs, the local ultraluminous infrared galaxy Arp220, and a local quiescent galaxy \citep[similar to the Milky Way;][]{chapman2003} to predict the 850$\micron$/3GHz flux ratio (Figure~\ref{fig:fig5-4_flux_ratio}).
The predicted flux ratio based on the average SED of SMGs flattens at higher redshifts. At $z >$ 4, the negative $k$-correction stabilizes the 850 $\micron$ flux, while the rest-frame frequency corresponding to the observed 3 GHz shifts to higher frequencies. Free-free emission begins to dominate the radio spectrum, causing the radio spectral index to flatten and the decrease in radio flux to slow, resulting in an overall flattening of the flux ratio. 
At even higher redshifts ($z >$ 7.5), the rest-frame wavelength corresponding to 850 $\micron$ approaches and then passes the peak of the thermal emission (assuming a dust temperature of 30 K, corresponding to $\sim$ 100 $\micron$ \citep{cy99}), leading to a gradual decline in the submillimeter flux. Consequently, the flux ratio becomes even flatter and even decline.

At low redshifts, SSA22 SMGs appear to show a tendency to be consistent with templates of quiescent galaxies to a moderate degree, while at higher redshifts they approach the hotter Arp220-like systems, despite evidence suggesting SMGs are not simple high-redshift analogs of local ULIRGs.
But, our results clearly demonstrate that no single template adequately describes the observed distribution. In contrast to the steep redshift evolution predicted by individual templates, the overall trend of the S$_{\text{850\micron}}$/S$_{\text{3GHz}}$ with redshift is relatively flat and can be described by: ${S_{\text{850\micron}}}/{S_{\text{3GHz}}} = (78.83 \pm 23.09) \times (1+z)^{(0.70 \pm 0.22)}$.

However, we caution that potential misidentification issues may exist. First, sources below redshift 1 could arise from incorrect cross-identification \citepalias{paperiii}. Second, since our sample identification relies on radio-band counterparts, we may inadvertently include some galaxies that are faint in the submillimeter but bright in the radio and mid-infrared bands, such as warm, radio-loud systems \citep{chapman2004}.

This discrepancy primarily arises from intrinsic diversity of galaxy types among SMGs \citep{chapman2005, casey2012}. \citet{grup2019} noted that infrared-selected samples encompass galaxies with varied SED shapes, with different shapes dominating infrared emission at different redshifts, making single-template approaches problematic. 
Even the average SMG SED fails to fully capture the observed scatter, indicating intrinsic SED diversity among SMGs. Furthermore, the average SED represents high-redshift DSFGs and is naturally less suitable for lower-redshift sources.
Additionally, the degeneracy between redshift and dust properties—such as varying dust temperatures and emissivities—prevents the submillimeter-to-radio flux ratio from uniquely determining redshift \citep{blain1999, chapman2005, casey2012}. As discussed previously, these galaxies may also exhibit dust temperatures that may evolve with redshift (Section~\ref{sec:sed_bin}), along with changes in the radio spectral shape at high redshifts (Section~\ref{sec:irrc_does_evolve}). Furthermore, factors such as AGN activity, dust mass and its spatial distribution can also introduce uncertainty.

Therefore, neither a single model nor the average SED of SMGs can tightly constrain the flux ratios of individual SMGs, leading to significant limitations and potentially poor estimation accuracy for individual source.
Nevertheless, due to the relatively concentrated redshift distribution of SMGs, the average SED of SMGs and ULIRGs template remain useful for estimating the mean redshift of a sample \citep{chapman2005, casey2012}.

\section{Summary}

This paper presents an analysis of the spectral energy distributions, star formation histories, and infrared-radio correlation for 221 850 \micron-selected SMGs in the SSA22 deep field (see \citetalias{paperiii}). 
The principal results are summarized as follows:

\begin{itemize}
\item[1.] 
We constructed the average SED for the sample of 221 SMGs (Figure~\ref{fig:fig3-1_average_sed}). Considerable variation is observed in the optical to near-infrared regimes among individual galaxies. The SED characteristics of SMGs differ from those of typical local ULIRGs or starburst systems, exhibiting cooler dust temperatures. This likely reflects more extended star formation and dust distributions, along with lower mid-infrared dust extinction in SMGs \citep{menendez2009, symeonidis2013}.

\item[2.] 
SED shapes also vary with evolutionary stage (Figure~\ref{fig:fig3-1_average_sed} panel (c)). Quiescent systems display cooler dust and older stellar populations, indicated by a more prominent 1.6 $\micron$ bump, stronger Balmer/4000$\rm \AA$ breaks, and rising far-UV flux due to the accumulation of low-mass post-AGB stars \citep{bc03}. In contrast, actively starbursting SMGs show stronger dust attenuation and warmer dust temperatures.

\item[3.] 
We provide average SEDs grouped by various physical and observational properties (Figure~\ref{fig:fig3-2_bin_sed}). With increasing redshift, SMGs exhibit higher total luminosity and warmer dust temperatures. Galaxies with greater dust mass tend to have lower infrared luminosity, colder dust, and higher submillimeter flux/luminosity. 
Higher $A_V$ correlates with stronger overall dust emission but does not systematically alter dust temperature, suggesting that attenuation is governed by multiple physical processes rather than a single mechanism. 
Because SMGs in different submillimeter flux bins exhibit similar SED shapes, the 850 $\micron$ flux alone is insufficient to distinguish between different galaxy types.

\item[4.] 
The majority of SMGs lie at the high-mass end of the star-forming main sequence, with starburst systems comprising $\sim$ 26\% of the sample (Figure~\ref{fig:fig4-3_ms}). We identify a “characteristic stellar mass” of $M_{\text{star}} \sim 10^{11}$ M$_\odot$, above which SMGs are predominantly on the main sequence or in the quenched phase (Figure~\ref{fig:fig4-4_deltams}). 
At a given evolutionary stage, SMGs at lower redshifts tend to have lower stellar masses than their high-redshift counterparts (Figure~\ref{fig:fig4-4_deltams}). 
The sSFR of SMGs decreases with decreasing redshift (Figure~\ref{fig:fig4-5_ssfr}), revealing a clear ``downsizing" effect: more massive galaxies complete their major mass assembly earlier and quench first, shifting the locus of star formation in the SMG population toward lower-mass systems at later times.

\item[5.] 
We investigated the star formation histories of SMGs (Figure~\ref{fig:fig4-1_spore}). 
The median mass-weighted age of the sample is 567 Myr, with substantial variation across different evolutionary stages. The majority of quenched SMGs have undergone extended evolutionary histories, averaging $\sim$ 2 Gyr. In contrast, rapidly evolving quenched systems typically experienced two intense, closely spaced starbursts, resulting in a much younger average age of $\sim$ 0.4 Gyr. 
Starbursting SMGs generally undergo a secondary star formation episode, which contributes significantly to their mass assembly—particularly for those with longer star formation histories.
We also identify a minority of massive galaxies that underwent extremely rapid mass assembly in the early universe, with star formation efficiencies reaching $\sim$ 0.2–0.8 (Section~\ref{sec:sfh_assembly}).

\item[6.] 
Using mass-weighted stellar ages and gas depletion timescales, we reconstruct the evolutionary timeline of 850 \micron–selected SMGs (Figure~\ref{fig:fig4-2_lifetime}). Most galaxies began forming stars around 1.68 Gyr after the Big Bang and underwent approximately 1 Gyr of evolution before being observed in their “SMG state”. They then exit the SMG phase within roughly 0.2 Gyr of active star formation, subsequently evolving into quiescent, redder systems (Section~\ref{sec:lifetime}).

\item[7.] 
Our SMG sample contributes, on average, $\sim$ 21\% to the cosmic SFRD (Figure~\ref{fig:fig4-6_sfrd}). This fraction peaks at 50-60\% in the redshift range $z =$ 2.5-3.5, but declines sharply at higher redshifts ($z =$ 4-6) due to changes in the star formation mode \citep{bourne2017, dunlop2017, bouwens2020, bouwens2022, zavala2021}. 
At lower redshifts, the contribution decreases further due to the ``downsizing" effect, declining baryonic conversion efficiency, and wavelength selection biases.
We also find that SMGs in overdense regions of the field make a significant contribution to the cosmic infrared background. 
The sample contributes $\sim$ 28\% to the cosmic SMD, with a redshift evolution trend consistent with its SFRD contribution (Figure~\ref{fig:fig4-7_smd}).
The number density of SMGs evolves rapidly from the early universe to $z \sim$ 1-3, increasing from $\sim$ 3-4$\times 10^{-6}$ cMpc$^{-3}$ to $\sim$ 4-6$\times 10^{-5}$ cMpc$^{-3}$ (Figure~\ref{fig:fig4-8_contribution}).
At lower redshifts ($z \lesssim$ 1-1.5), 850\,$\micron$ surveys miss a population of submillimeter-faint galaxies with warmer dust temperatures due to gas exhaustion, downsizing, and bandpass selection effects (Section~\ref{sec:sfh_evolution}).

\item[8.] 
The average infrared–radio correlation parameter $q_\text{IR}$ for our sample is 2.37, with a mild evolution observed as $q_\text{IR} \propto (1+z)^{-0.11}$ (Figure~\ref{fig:fig5-2_qir}). 
This trend may be influenced by the presence of AGNs at high redshifts. Additionally, the evolution in $q_\text{IR}$ also arises from intrinsic differences between low- and high-redshift SMGs, potentially linked to the ``downsizing" effect or the increasing similarity of low-redshift SMGs to local star-forming galaxies (Section~\ref{sec:irrc_q}, Section~\ref{sec:irrc_does_evolve}).
The submillimeter-to-radio flux ratio evolves gently as $S_{\text{850\micron}}/S_{\text{3GHz}} \propto (1+z)^{0.70}$, and no single SED template adequately captures its distribution—highlighting the intrinsic diversity in SMG spectral shapes (Figure~\ref{fig:fig5-4_flux_ratio}).

\end{itemize}


\begin{acknowledgments}
\section*{acknowledgments}
We sincerely thank Ian Smail for his constructive comments and insightful suggestions, which significantly improved the rigor and clarity of this manuscript.

Y.A. acknowledges the support from the National Key R\&D Program of China (2023YFA1608204), the National Natural Science Foundation of China (NSFC grant 12173089), the China Manned Space Program with grant no. CMS-CSST-2025-A10, and  the Strategic Priority Research Program of the Chinese Academy of Sciences (grant No.XDB0800301).

K.M. acknowledges the Waseda University Grant for Special Research Projects (Project number: 2025C-484).

The James Clerk Maxwell Telescope is operated by the East Asian Observatory on behalf of The National Astronomical Observatory of Japan; Academia Sinica Institute of Astronomy and Astrophysics; the Korea Astronomy and Space Science Institute; the National Astronomical Research Institute of Thailand; Center for Astronomical Mega-Science (as well as the National Key R\&D Program of China with No. 2017YFA0402700). Additional funding support is provided by the Science and Technology Facilities Council of the United Kingdom and participating universities and organizations in the United Kingdom and Canada. Additional funds for the construction of SCUBA-2 were provided by the Canada Foundation for Innovation.

\end{acknowledgments}

%



\vspace{5mm}
\software {astropy \citep{astropy2013, astropy2018, astropy2022}, pandas \citep{reback2020pandas}, numpy \citep{harris2020array}, matplotlib \citep{Hunter2007}, scipy \citep{2020SciPy-NMeth,mckinney-proc-scipy-2010}, }



\appendix


\bibliography{atext, other}

\bibliographystyle{aasjournal}



\end{document}